\documentclass[preprint,aps, amsmath,amssymb]{revtex4-1}
\newcommand{\be}{\begin{equation}}
\newcommand{\ee}{\end{equation}}
\newcommand{\bea}{\begin{eqnarray}}
\newcommand{\eea}{\end{eqnarray}}
\newcommand{\ba}{\begin{array}}
\newcommand{\ea}{\end{array}}

\usepackage{MnSymbol}
\usepackage[utf8]{inputenc}
\usepackage{soulutf8}
\usepackage{graphicx}
\usepackage{epstopdf}
\usepackage{amsmath}
\usepackage{slashed}
\usepackage{mathtools}
\usepackage{slashed}
\usepackage{array,multirow}
\usepackage{booktabs}

\usepackage{blindtext}

\begin{document}

\title{Light-front holographic radiative transition form factors for light mesons}

\author{Mohammad Ahmady}
\email{mahmady@mta.ca}
\affiliation{\small Department of Physics, Mount Allison University, Sackville, New Brunswick, Canada, E4L 1E6}

\author{Satvir Kaur}
\email{satvirkaur578@gmail.com}
\affiliation{\small Department of Physics, Dr. B. R. Ambedkar National Institute of Technology, Jalandhar 144011, India}

\author{Chandan Mondal}
\email{mondal@impcas.ac.cn}
\affiliation{\small Institute of Modern Physics, Chinese Academy of Sciences, Lanzhou 730000, China $\&$ School of Nuclear Science and Technology, University of Chinese Academy of Sciences, Beijing 100049, China}

\author{Ruben Sandapen}
\email{ruben.sandapen@acadiau.ca}
\affiliation{\small Department of Physics, Acadia University, Wolfville, Nova-Scotia, Canada, B4P 2R6}

\begin{abstract} 
We predict the $\mathcal{V} \to \mathcal{P} \gamma$ decay widths and the $\mathcal{V} \to \mathcal{P} \gamma^{*}$ transition form factors, where $\mathcal{V}=(\rho, \omega, K^*, \phi)$ and  $\mathcal{P}= (\pi,K, \eta,\eta^\prime)$,  using spin-improved holographic light-front wavefunctions for the mesons. We find excellent agreement with the available data for both the decay widths and the timelike transition form factors extracted from the leptonic conversion decays $\mathcal{V} \to \mathcal{P} l^+ l^-$. 
\end{abstract}

\maketitle
         
\section{Introduction}
\label{Sec:Introduction}
The  vector-to-pseudoscalar meson radiative transitions, $\mathcal{V} \to \mathcal{P} \gamma^{(*)}$, are important probes of quark confinement dynamics, encoded in their transition form factors at zero or low momentum transfer. These non-perturbative form factors are universal and appear in other physical processes like the hadronic light-by-light contribution to the Standard Model prediction of the muon anomalous magnetic moment \cite{Jegerlehner:2009ry}. On the experimental side, there exists measurements of the  $\mathcal{V} \to \mathcal{P} \gamma$ decay widths \cite{PDG} and of the $\mathcal{V} \to \mathcal{P} \gamma^*$  transition form factors for low-momentum timelike photons. The latter are extracted from the leptonic conversion decays $\mathcal{V} \to \mathcal{P} l^+ l^-$: $\omega \to \pi^0 \mu^+ \mu^-$ in the Lepton-G and NA60 experiments \cite{Dzhelyadin:1980tj,Arnaldi:2009aa,Arnaldi:2016pzu}; $\omega \to \pi^0 e^+ e^-$ in the A2MM experiment \cite{Adlarson:2016hpp} and $\phi \to \eta e^+ e^-$ in the SND and KLOE experiments \cite{Achasov:2000ne, Babusci:2014ldz}. The discrepancy between the Lepton-G and NA60 data with the prediction of the Vector Meson Dominance (VMD) model has triggered considerable theoretical attention \cite{TERSCHLUSEN2010191,Schneider:2012ez,Ivashyn:2011hb,PhysRevD.91.094029}, and prompted the measurement of the $\phi \to \pi^0 e^+ e^-$ decay by the KLOE experiment \cite{Anastasi:2016qga}. The transition form factors have also been predicted using the Dyson-Schwinger Equations \cite{Maris:2002mz}, the pQCD factorization approach \cite{Zhang:2015mxa}, chiral perturbation theory \cite{Kimura:2016xnx,Feldmann:1999uf}, and,  more relevant to this paper, in the light-front formalism \cite{Choi:1996mq,Choi:2008si,Yu_2007}, where they are expressed as overlap integrals of the meson light-front wavefunctions.

Here, we compute the transition form factors using spin-improved holographic light-front wavefunctions for the mesons. These spin-improved wavefunctions  were proposed for the vector mesons $\mathcal{V}=(\rho,K^*,\phi)$ in Refs. \cite{Forshaw:2012im,Ahmady:2013cva,Ahmady:2016ujw} and for the pseudoscalar mesons $\mathcal{P}=(\pi,K, \eta, \eta^\prime)$ in Refs. \cite{Ahmady:2016ufq,Ahmady:2018muv}. The vector meson wavefunctions were used to predict their decay constants, the cross-sections for diffractive $(\rho/\phi)$-electroproduction and several observables for the semileptonic $B_{(s)} \to (\rho,K^*,\phi) + l \bar{l}$ decays \cite{Ahmady:2014cpa,Ahmady:2015fha,Ahmady:2018fvo,Ahmady:2019hag,Ahmady:2013cva}.  The pseudoscalar meson wavefunctions were used to predict their decay constants, electromagnetic elastic form factors and charge radii \cite{Ahmady:2016ufq, Ahmady:2018muv}, as well as the $(\pi^0,\eta, \eta^\prime)  \to \gamma \gamma^{*}$ transition form factors and $(\pi^0,\eta, \eta^\prime)  \to \gamma \gamma$ decay widths \cite{Ahmady:2018muv}. So far, the spin-improved holographic wavefunctions have been used in  processes involving only one light (pseudoscalar or vector) meson. Our goal in this paper is to use them simultaneously to predict the $\mathcal{V} \to \mathcal{P} \gamma^{*}$ transition form factors and the corresponding $\mathcal{V} \to \mathcal{P} \gamma$ decay widths. For completeness, we shall also predict the measured $\eta^\prime \to (\rho,\omega) \gamma$ decay widths. 

In our approach, the difference between pseudoscalar and vector mesons lies in the quark-antiquark helicity wavefunction that modifies their universal holographic wavefunction. The dynamical part of the latter satisfies the holographic Schr\"odinger Equation

\begin{equation}
	\left(-\frac{\mathrm{d}^2}{\mathrm{d}\zeta^2}-\frac{4L^2-1}{4\zeta^2} + U(\zeta) \right)\phi(\zeta)=M^2 \phi(\zeta) \;,
\label{hSE-mesons}
\end{equation} 
where $\zeta=\sqrt{x(1-x)} b_\perp$, with $x=k^+/P^+$ being the light-front momentum fraction carried by the quark and $b_\perp$ is the magnitude of the transverse separation, $\mathbf{b}=b_\perp e^{i\theta_\perp}$, between the quark and antiquark. Eq. \eqref{hSE-mesons} can be derived in light-front QCD in a semiclassical approximation where quark masses and quantum loops are neglected \cite{Brodsky:2006uqa,deTeramond:2005su,Brodsky:2014yha,deTeramond:2008ht}. More interestingly, mapping $\zeta$ onto the fifth dimension, $z$, in anti-de Sitter spacetime, $\mathrm{AdS}_5$, Eq. \eqref{hSE-mesons} becomes the wave equation for the amplitude of spin-$J$ string modes propagating in a modified $\mathrm{AdS}_5$ spacetime, where $(2-J)^2=L^2 -(\mu R)^2$ with $\mu$ being the $5$-d mass of the string modes and $R$ the radius of curvature of $\mathrm{AdS}_5$ \cite{Brodsky:2014yha}. The geometry of  $\mathrm{AdS}_5$ is distorted by a dilaton field $\varphi(z)$ which drives the confining potential in physical spacetime:
\begin{equation}
	U(\zeta, J)= \frac{1}{2} \varphi^{\prime\prime}(z) + \frac{1}{4} \varphi^{\prime}(z)^2 + \left(\frac{2J-3}{2 z} \right)\varphi^{\prime} (z) 
\label{dilation-U}
\end{equation}
with $\zeta \leftrightarrow z$. While Eq. \eqref{dilation-U} is true for an arbitrary dilaton field, only a quadratic confinement potential, $U(\zeta)=\kappa^4 \zeta^2$, leaves the underlying action leading to Eq. \eqref{hSE-mesons} conformally invariant \cite{Brodsky:2013ar}, and this, in turn, requires the dilaton field to be also quadratic, $\varphi=\kappa^2 z^2$. Then, Eq. \eqref{dilation-U} yields
\begin{equation}
	U(\zeta, J)=\kappa^4 \zeta^2 + 2 \kappa^2(J-1) \;.
\end{equation}
The mass scale $\kappa$ which simultaneously sets the strength of the dilaton field in $\mathrm{AdS}_5$ and the hadron mass scale in physical spacetime, is referred to as the AdS/QCD mass scale.

The supersymmetrization  of Eq. \eqref{hSE-mesons}  leads to the identification of mesons and baryons (considered as quark-diquark systems) as supersymmetric partners, provided that they differ by only one unit of orbital angular momentum \cite{Dosch:2015nwa,Nielsen:2018uyn,Brodsky:2020ajy}. In other words, the meson and baryon mass spectra are given by
\begin{equation}
	M_M^2=4 \kappa^2(n+L) + 2\kappa^2 S \hspace{1cm}\mathrm{and}\hspace{1cm} M_B^2=4 \kappa^2 (n + L +1) + 2\kappa^2 S
	\label{meson-masses}
\end{equation}
where $S$ is the spin of the quark-antiquark in mesons and the lowest possible value of the diquark spin in baryons. The lightest hadron (with $J=L=S=0$) is massless, with no supersymmetric partner, and is naturally identified with the pion. At this point, the only free parameter is the mass scale, $\kappa$, and it can be fixed by a simultaneous fit to the Regge slopes of light mesons and baryons. This fit yields $\kappa=523 \pm 24$ MeV \cite{Brodsky:2016rvj}, which we refer to as the universal AdS/QCD mass scale.

Solving Eq. \eqref{hSE-mesons} yields the dynamical part of the holographic meson wavefunction,
\begin{equation}
 	\phi_{nL}(\zeta)= \kappa^{1+L} \sqrt{\frac{2 n !}{(n+L)!}} \zeta^{1/2+L} \exp{\left(\frac{-\kappa^2 \zeta^2}{2}\right)}  ~ L_n^L(\kappa^2 \zeta^2) 
 \label{phi-zeta}
 \end{equation}
 and the complete meson wavefunction is given by \cite{Brodsky:2014yha}
\begin{equation}
	\Psi_{nL} (x,\zeta,\varphi)=\frac{\phi_{nL}(\zeta)}{\sqrt{2\pi \zeta}} X(x) e^{iL\varphi} \;,
	\end{equation}
where $X(x)$ is fixed by mapping the spacelike electromagnetic form factor of the pion in $\mathrm{AdS}_5$ and in physical spacetime \cite{Brodsky:2008pf}. In $\mathrm{AdS}_5$, the form factor is given by an overlap integral of the ingoing and outgoing hadronic modes convoluted with the bulk-to-boundary propagator which maps onto the free electromagnetic current in physical spacetime. In physical spacetime, the form factor is given by an integral overlap of the meson light-front wavefunctions, i.e. the Drell-Yan-West formula \cite{Drell:1969km,West:1970av}. This procedure yields $X(x)=\sqrt{x(1-x)}$ \cite{Brodsky:2014yha}. Matching of the $\mathrm{AdS}_5$ and physical spacetime gravitational form factors gives an identical result \cite{Brodsky:2008pf}.  

The normalized holographic light-front wavefunction for mesons with $n=L=0$ is given by
 \begin{equation}
 	\Psi (x,\zeta^2) = \frac{\kappa}{\sqrt{\pi}} \sqrt{x \bar{x}}  \exp{ \left[ -{ \kappa^2 \zeta^2  \over 2} \right] } 
\label{pionhwf} 
\end{equation}
or, in momentum space,
\begin{equation}
	\Psi (x, k_\perp^2) \propto \frac{1}{\sqrt{x\bar{x}}} \exp \left( -\frac{M^2}{2\kappa^2} \right)
\label{hwf-k}
\end{equation}
where $M^2= k_\perp^2/x\bar{x}$ is the invariant mass of the quark-antiquark pair. Here $k_\perp$ is the magnitude of the two-dimensional transverse momentum $\mathbf{k}=k_\perp e^{i\theta_{k_\perp}}$ which is the Fourier conjugate of the transverse distance $\mathbf{b}$ between the quark and the antiquark. For non-zero quark masses,  this invariant mass should be $M_{f\bar{f}^\prime}^2= (k_\perp^2 + \bar{x} m_f^2 + x m^2_{\bar{f}^\prime} )/x\bar{x}$, where $f$ and $\bar{f}^\prime$ denote the flavours of the quark and antiquark respectively. This motivates a prescription \cite{Brodsky:2008pg} to account for light quark masses: replace $M^2$ by $M^2_{f\bar{f}^\prime}$ in Eq. \eqref{hwf-k}. Then, the holographic wavefunction becomes
\begin{equation}
	\Psi (x, k_\perp^2) \propto \frac{1}{\sqrt{x\bar{x}}} \exp \left( -\frac{k_\perp^2}{2\kappa^2 x\bar{x}} \right) \exp \left( -\frac{1}{2\kappa^2}\left(\frac{m_f^2}{x} + \frac{m_{\bar{f}^\prime}^2}{\bar{x}}\right)\right) \;.
	\label{hwf-k-mf}
\end{equation}
So far, the quark and antiquark helicity indices have been suppressed \cite{Brodsky:2014yha}. Making them explicit, we have
\begin{equation}
	\Psi^{\mathcal{P},\mathcal{V}}_{h, \bar{h}}(x,\mathbf{k}) = S^{\mathcal{P},\mathcal{V}}_{h,\bar{h}} \Psi(x, k_\perp^2)  \;,
\label{non-dynamical-spin}
\end{equation}
where 
\begin{equation}
	S^\mathcal{P}_{h,\bar{h}} = \frac{1}{\sqrt{2}} h \delta_{h,-\bar{h}} 
\label{non-dynamical-spin-P}
\end{equation}
and
\begin{equation}
	S^{\mathcal{V}(L)}_{h,\bar{h}} = \frac{1}{\sqrt{2}} \delta_{h,-\bar{h}} \hspace{1cm};\hspace{1cm} S^{\mathcal{V}(T)}_{h,\bar{h}} = \frac{1}{\sqrt{2}} \delta_{h \pm,\bar{h}\pm} \;.
\label{non-dynamical-spin-V}
\end{equation}
With a universal AdS/QCD scale, this would lead to degenerate decay constants for the pseudoscalar and vector mesons, as well as degenerate decay constants for the longitudinally and transversely polarized vector mesons, in contradiction with experiment \cite{PDG} and lattice QCD  \cite{Becirevic_2003,PhysRevD.68.054501}. Indeed, in light-front holography, there is no distinction between the dynamical wavefunctions of ground state ($n=L=0$) light pseudoscalar and vector mesons since Eq. \eqref{phi-zeta} does not depend on $S$.

\section{Dynamical spin effects} 
\label{Sec:Spin}
The above shortcomings can be addressed by taking into account dynamical spin effects. The pseudoscalar and vector meson wavefunctions are then given by \cite{Forshaw:2012im,Ahmady:2013cva,Ahmady:2016ufq,Ahmady:2018muv}
\begin{equation}
	\Psi^{\mathcal{P},\mathcal{V}}_{h,\bar{h}} (x, \mathbf{k})=   S^{\mathcal{P},\mathcal{V}}_{h, \bar{h}}(x, \mathbf{k}) \Psi(x, k_\perp^2)\;,
\label{spin-improved-wf}
\end{equation}
where $\Psi(x, k_\perp^2)$ is the holographic wavefunction given by Eq. \eqref{hwf-k-mf}, and the Lorentz invariant spin structures are given by
 \begin{equation}
	S^{\mathcal{P},\mathcal{V}}_{h, \bar{h}}(x, \mathbf{k})= \frac{\bar{u}_{h}(xP^+,\mathbf{k})}{\sqrt{x}} \Gamma_{\mathcal{P},\mathcal{V}} \frac{v_{\bar{h}}(\bar{x}P^+,-\mathbf{k})}{\sqrt{\bar{x}}} 
\label{spin}
\end{equation}
with 
\begin{equation}
\Gamma_\mathcal{V}=\varepsilon^{\lambda}_\mathcal{V} \cdot \gamma 
\label{spin-structure-V}
\end{equation}
where
\begin{equation}
\varepsilon^L_\mathcal{V}= 	\left(\frac{P^+}{M_{\mathcal{V}}}, -\frac{M_{\mathcal{V}}}{P^+}, 0, 0\right) \hspace{1cm};\hspace{1cm} \varepsilon^{T(\pm)}_{\mathcal{V}}=\mp \frac{1}{\sqrt{2}}(0,0,1, \pm i) 
\end{equation}
and
\begin{equation}
\Gamma_{\mathcal{P}} =  (P \cdot \gamma) \gamma^5 +   M_{\mathcal{P}} \gamma^5 \;,
\label{spin-structure-P}
\end{equation}
where
\begin{equation}
	P^\mu=\left(P^+, \frac{M^2_{\mathcal{P}}}{P^+}, 0,0\right) \;.
\end{equation}

Eq. \eqref{spin-structure-V} is modelled upon the photon-quark-antiquark vertex and leads to a successful description of diffractive $\rho$ and $\phi$ electroproduction \cite{Forshaw:2012im,Ahmady:2016ujw}. On the other hand, Eq. \eqref{spin-structure-P} does not give a good description of the pseudoscalar meson data. However, since the individual terms of Eq. \eqref{spin-structure-P} are separately Lorentz invariant, we are able to use the more flexible structure,
\begin{equation}
 \Gamma_{\mathcal{P}} =  \frac{M^2_\mathcal{P}}{2P^+} \gamma^+ \gamma^5 + \frac{A P^+}{2} \gamma^- \gamma^5 + B M_{\mathcal{P}} \gamma^5  
 \label{spin-structures-ABC}
\end{equation}
where $A$ and $B$ are dimensionless constants which quantify the importance of dynamical spin effects. Indeed, setting $A=B=0$, we are left with the non-dynamical $\gamma^+\gamma^5$ spin structure which yields Eq. \eqref{non-dynamical-spin-P}. Refs. \cite{Ahmady:2018muv,Ahmady:2016ufq} choose $A=0$, as required by the data, while the situation is less clear for $B$: the pion data favour $B \ge 1$, the (charged) kaon data prefer $B=0$. For the $\eta/\eta^\prime$ system, the $\eta/\eta^\prime \to \gamma \gamma^*$ transition form factor data prefer $B \gg 1$ while the $\eta (\eta^\prime) \to \gamma \gamma$ decay widths data prefer $B=0~(B=1)$. Consequently, we are compelled to treat $B$ as a free parameter here. 

Explicitly, the spin-improved holographic wavefunctions are given by \cite{Ahmady:2018muv,Ahmady:2016ufq}
\begin{equation}
	\Psi^{\mathcal{P}}_{h,\bar{h}}(x,\mathbf{k})=\mathcal{N} \left[\left(M_{\mathcal{P}}  +   B \left(\frac{m_f}{x} + \frac{m_{\bar{f}}}{\bar{x}}\right)\right) h \delta_{h,-\bar{h}}  - B \left(\frac{k_\perp e^{-ih\theta_{k_\perp}}}{x\bar{x}}\right) \delta_{h,\bar{h}}\right] \Psi(x, k_\perp^2) \;,
\label{pseudoscalar}
\end{equation}
while \cite{Forshaw:2012im,Ahmady:2013cva}
\begin{equation}
	\Psi_{h,\bar{h}}^{\mathcal{V}(L)}(x,\mathbf{k})= \mathcal{N}_{L} \delta_{h,-\bar{h}} \left(M_\mathcal{V}^2  +  \left(\frac{m_f m_{\bar{f}} + k_\perp^2}{x\bar{x}} \right)\right) \Psi (x, k_\perp^2)
\label{vector-L}
\end{equation}
and
\begin{equation}
	\Psi_{h,\bar{h}}^{\mathcal{V}(T=\pm)}(x,\mathbf{k})=\frac{\mathcal{N}_{T}}{\sqrt{2}} \left[\pm k_\perp e^{\pm i \theta_{k_\perp}} \left(\frac{\delta_{h,\pm} \delta_{\bar{h},\mp}}{\bar{x}}-\frac{\delta_{h,\mp} \delta_{\bar{h},\pm}}{x}\right)  + \left(\frac{m_f}{x} + \frac{m_{\bar{f}}}{\bar{x}}\right) \delta_{h,\pm} \delta_{\bar{h},\pm}\right] \Psi(x, k_\perp^2) \;.
	\label{vector-T} 
	\end{equation}
The normalization constants $\mathcal{N}_{(L,T)}$ are fixed using
\begin{equation}
	 \sum_{h,\bar{h}} \int \frac{\mathrm{d}^2 \mathbf{k}}{16\pi^3} \mathrm{d} x |\Psi_{h,\bar{h}}^{\mathcal{P},\mathcal{V}}(x,\mathbf{k}) |^2=1 \;,   
	 \end{equation}
which embodies the assumption that the meson consists only of a quark-antiquark pair. Alternative spin-improved holographic wavefunctions have been proposed in Refs. \cite{Chang:2016ouf,Chang:2018aut}.

We remark that our spin-improved holographic light-front wavefunctions are distinct from the so-called ``boosted" wavefunctions obtained by ``boosting" the non-relativistic Schr\"odinger wavefunction in the meson's rest frame to the light-front. This is usually performed using the Brodsky-Huang-Lepage prescription \cite{Brodsky:1981jv}, together with the Melosh rotation \cite{PhysRevD.9.1095} for the spin structure: see, for example, Ref. \cite{Huang:1994dy}.  Our spin structures, Eq. \eqref{spin}, are fixed by the rules of light-front field theory for coupling a quark and an antiquark into a (point-like) meson while non-perturbative bound state effects are captured by the holographic wavefunction given by Eq. \eqref{hwf-k-mf}. Our wavefunctions are directly formulated on the light-front and are frame-independent, avoiding the ambiguities associated with a boosting prescription.  

Having said that, it is worth noting that the boosting of a harmonic oscillator rest frame Schr\"odinger wavefunction results is the boosted Gaussian wavefunction \cite{Forshaw:2003ki} which is similar to the holographic Gaussian wavefunction given by Eq. \eqref{hwf-k-mf}. However, we must highlight three essential differences between these two wavefunctions: first, as we mentioned before, the harmonic potential in the holographic Schr\"odinger Equation, Eq. \eqref{hSE-mesons}, is uniquely fixed by a specific mechanism of conformal symmetry breaking \cite{Brodsky:2013ar} in semiclassical light-front QCD  unlike the assumed harmonic potential in the ordinary Schr\"odinger Equation. Second the AdS/QCD mass scale, $\kappa$, is extracted from spectroscopic data and it fixes the width of the holographic Gaussian where as the width of the boosted Gaussian is a free parameter which has to be fixed by some constraint on the wavefunction \cite{Huang:1994dy,Forshaw:2003ki}.  Thirdly, the two wavefunctions differ by an overall factor of $1/\sqrt{x\bar{x}}$  which makes their end-point behaviours different. Interestingly, the data on diffractive $\rho$ meson electroproduction are able to discriminate between the two wavefunctions and favour the holographic Gaussian \cite{Forshaw:2012im}.

\section{Radiative Transition Form Factors}
\label{Sec:TFF}

\begin{figure}[hbt]
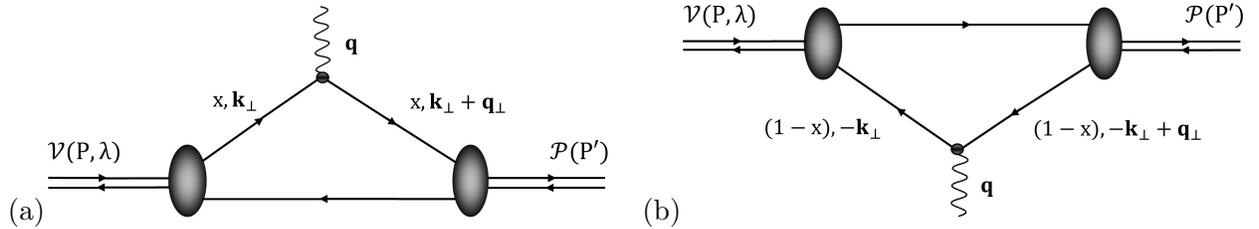

(a)\includegraphics[width=7.5cm]{Active-quark.pdf} \hspace{0.2cm}
(b)\includegraphics[width=7.5cm]{Active-antiquark.pdf}   \caption{The active quark contribution (a) and the active antiquark contribution (b) to the $\mathcal{V} \to \mathcal{P} \gamma^*$ transition.}
   \label{Fig:Feynman-diagrams}  
   \end{figure}

The transition form factors, $F_{\mathcal{V}\mathcal{P}}(Q^2)$, are defined by \cite{Choi:1996mq}
\begin{equation}
	i F_{\mathcal{V}\mathcal{P}}(q^2) \epsilon^{\mu \nu \rho \sigma} \varepsilon_{\nu}^\lambda P_\rho^\prime P_\sigma = \langle \mathcal{P}(P^\prime) | J_{\mathrm{em}}^\mu(0) | \mathcal{V} (P, \lambda) \rangle \;,
\label{TFF-def}
\end{equation}
where $P (P^\prime)$ is the $4$-momentum of the vector (pseudoscalar) meson, $q^2=(P^\prime-P)^2$ is the spacelike $4$-momentum transfer, and $J_{\mathrm{em}}^\mu(0)$ is the quark electromagnetic current. To leading order in $\alpha_{\textrm{em}}$, there are two contributions to the radiative transition matrix element, with the photon being either radiated by the quark or the antiquark, as shown in Fig. \ref{Fig:Feynman-diagrams}. Focusing on states with a specified flavour content, we can write:
\begin{equation}
\langle \mathcal{P}; f\bar{f}^\prime | J_{\mathrm{em}}^\mu(0) | \mathcal{V}; f\bar{f}^\prime  \rangle=\langle \mathcal{P}; f\bar{f}^\prime| J_{f}^\mu(0) | \mathcal{V};f\bar{f}^\prime\rangle + \langle \mathcal{P};f\bar{f}^\prime| J_{f^\prime}^\mu(0) | \mathcal{V};f\bar{f}^\prime\rangle
\end{equation}
with
\begin{equation}
	J^\mu_{f}(0)= e_f \int \frac{\mathrm{d} k^+ \mathrm{d}^2 \mathbf{k}}{16 \pi^3 k^+} \frac{\mathrm{d} k^{\prime+} \mathrm{d}^2 \mathbf{k^\prime}}{16 \pi^3 k^{\prime +}}   \hat{b}_f^\dagger(k^+,\mathbf{k}) \hat{b}_f(k^{\prime +},\mathbf{k}^\prime) \bar{u}_f(k^+,\mathbf{k})\gamma^\mu u_f(k^{\prime +},\mathbf{k}^\prime)
\label{Jq}
\end{equation}
and
\begin{equation}
	J^\mu_{{\bar{f}^\prime}}(0)= e_{\bar{f}^\prime} \int\frac{\mathrm{d} k^+ \mathrm{d}^2 \mathbf{k}}{16 \pi^3 k^+} \frac{\mathrm{d} k^{\prime+} \mathrm{d}^2 \mathbf{k^\prime}}{16 \pi^3 k^{\prime +}} \hat{d}_{\bar{f}^\prime}^{\dagger}(k^+,\mathbf{k}) \hat{d}_{\bar{f}^\prime}(k^{\prime +},\mathbf{k}^\prime) \bar{v}_{\bar{f}^\prime}(k^+,\mathbf{k})\gamma^\mu v_{\bar{f}^\prime}(k^{\prime +},\mathbf{k}^\prime)
\label{Jbarq}
\end{equation}	
where, for notational simplicity, we have suppressed the helicity and colour indices. For the non-strange mesons, $\mathcal{P}=(\pi, \eta, \eta^\prime)$ and $\mathcal{V}=(\rho,\omega, \phi)$, Eq. \eqref{Jq} and Eq. \eqref{Jbarq} map onto each other under a $G$-parity transformation, i.e. $J_{f}^\mu(0)=\hat{G} J_{\bar{f}^\prime}^\mu(0) \hat{G}^\dagger$, so that 
\begin{equation}
	\langle \mathcal{P};f\bar{f}^\prime | J_{f}^\mu(0) | \mathcal{V}; f\bar{f}^\prime \rangle = G_\mathcal{P} G_{\mathcal{V}}  (-1)^{I_\mathcal{P}} (-1)^{I_\mathcal{V}}\langle \mathcal{P}; f\bar{f}^\prime| J_{\bar{f}^\prime}^\mu(0) | \mathcal{V}; f\bar{f}^\prime\rangle \;,	
\end{equation}
where $G_\mathcal{P,V}$ and $I_\mathcal{P,V}$ are the G-parity and isospin quantum numbers. For the non-strange mesons, the $I^G$ assignments are: $\pi^{0,\pm} (1^-)$, $\rho^{0,\pm} (1^+)$, $\eta/\eta^\prime (0^+)$, and $\phi/\omega (0^-)$, implying that
	\begin{equation}
	\langle \mathcal{P};f\bar{f}^\prime | J_{f}^\mu(0) | \mathcal{V} ;f\bar{f}^\prime \rangle =-\langle \mathcal{P};f\bar{f}^\prime | J_{\bar{f}^\prime}^\mu(0) | \mathcal{V};f\bar{f}^\prime \rangle \;,
	\label{minus-relation-Feynman-graphs}
	\end{equation}
i.e. the two Feynman graphs of Fig. \ref{Fig:Feynman-diagrams} differ only by a minus sign. This is not the case for transitions involving the strange mesons.     

To proceed, we choose the ``good" current, $J_{\mathrm{em}}^+(0)$, in the Drell-Yan-West frame \cite{Drell:1969km,West:1970av} where
\begin{equation}
	P^\mu=\left(P^+, \frac{M^2_\mathcal{V}}{P^+}, \mathbf{0}\right) \hspace{1cm};\hspace{1cm} P^{\mu \prime}=\left(P^+, \frac{M^2_\mathcal{P}+ q_\perp^2}{P^+}, \mathbf{q}\right)
\end{equation}
i.e. with $q^+=0$ and $q^2=-q_\perp^2 <0$. This choice avoids the zero-mode contributions \cite{PhysRevD.72.013004} but, at the same time, restricts the computation of the transition form factor to the spacelike region where $Q^2 \equiv -q^2 >0$. However, it is possible to analytically continue the spacelike form factor to the timelike region using the prescription $q_\perp \to i q_\perp$ \cite{Melikhov:1995xz,Choi:1999bg,Yu_2007}. Note that the ``good" current matrix element vanishes for $\lambda=L$, and therefore we must take $\lambda=T$ (here we choose $T=+$) in order to extract $F_{\mathcal{V} \mathcal{P}}(Q^2)$. Fock expanding the meson states and using Eqs. \eqref{Jq} and \eqref{Jbarq}, we find that
\begin{equation}
\langle \mathcal{P}(P^\prime) | J_{\mathrm{em}}^+(0) | \mathcal{V} (P, +) \rangle = 2P^+  \sumint \frac{\mathrm{d} x \mathrm{d}^2 \mathbf{k}}{16\pi^3}  [e_f \Psi_{h \bar{h}}^{\mathcal{P}*}(x, \mathbf{k} + \bar{x} \mathbf{q}) + e_{\bar{f}^\prime}\Psi_{h \bar{h}}^{\mathcal{P}*}(x, \mathbf{k} - x \mathbf{q})]\Psi_{h \bar{h}}^{\mathcal{V}(+)} (x,\mathbf{k}) 
\label{overlap}
\end{equation}
where we have used the shorthand notation, $\sumint \equiv \sum_{h,\bar{h}} \int$.  Our derivation of Eq. \eqref{overlap} is standard and model-independent. We now use our spin-improved holographic wavefunctions, given by Eqs. \eqref{pseudoscalar} and \eqref{vector-T}, we obtain
\begin{equation}
  \sumint \frac{\mathrm{d}^2 \mathbf{k}}{16\pi^3}  \Psi_{h \bar{h}}^{\mathcal{P}*}(x, \mathbf{k} + \bar{x}  \mathbf{q}) \Psi_{h \bar{h}}^{\mathcal{V}(+)} (x,\mathbf{k}) =- \tilde{\mathcal{N}} \bar{x}\mathbf{q}   \exp{\left(-\frac{(q_\perp \bar{x})^2 + 4 (\bar{x} m_f^2 + x m^2_{\bar{f}^\prime})}{4 \kappa^2 x\bar{x}}\right)} \mathcal{M}(x)
  \label{active-quark}
  \end{equation}
and
\begin{equation}
  \sumint \frac{\mathrm{d}^2 \mathbf{k}}{16\pi^3}  \Psi_{h \bar{h}}^{\mathcal{P}*}(x, \mathbf{k} - x \mathbf{q}) \Psi_{h \bar{h}}^{\mathcal{V}(+)} (x,\mathbf{k}) = \tilde{\mathcal{N}} x \mathbf{q}  \exp{\left(-\frac{(q_\perp x)^2 + 4 (\bar{x} m_f^2 + x m^2_{\bar{f}^\prime})}{4 \kappa^2 x\bar{x}}\right)} \mathcal{M}(x) \;,
  \label{active-antiquark}
  \end{equation}
where $\tilde{\mathcal{N}} \equiv \mathcal{N} \mathcal{N}_T \kappa^2/(8\pi^2)$	 and
\begin{equation}
\mathcal{M}(x)= \frac{1}{x\bar{x}} \left( \frac{M_{\mathcal{P}}}{2} + B \left(\frac{\bar{x} m_f+ x m_{\bar{f}}}{x\bar{x}}\right) \right) \;.
\end{equation}
Inserting Eqs. \eqref{active-quark} and \eqref{active-antiquark} in Eq. \eqref{overlap}, Eq. \eqref{TFF-def} leads to
\begin{equation}
	F_{\mathcal{V} \mathcal{P}}(Q^2) = e_f I (q_\perp^2; M_\mathcal{P}, m_f, m_{\bar{f}^\prime}) - e_{\bar{f}^\prime} I (q_\perp^2; M_\mathcal{P}, m_{\bar{f}^\prime}, m_f) \;,
\label{TFF-int}
\end{equation}
where
\begin{equation}
	I (q_\perp^2; M_\mathcal{P}, m_f, m_{\bar{f}^\prime}) = \tilde{\mathcal{N}} \int \frac{\mathrm{d} x}{x} \left( \frac{M_{\mathcal{P}}}{2} + B \left(\frac{\bar{x} m_f+ x m_{\bar{f}^\prime}}{x\bar{x}}\right) \right)\exp{\left(-\frac{(q_\perp \bar{x})^2 + 4 (\bar{x} m_f^2 + x m^2_{\bar{f}^\prime})}{4 \kappa^2 x\bar{x}}\right)} \;.
\label{integral}
\end{equation}

For the non-strange mesons, $m_f=m_{\bar{f}^\prime}$, the two integrals in Eq. \eqref{TFF-int} are identical, i.e.
\begin{equation}
F_{\mathcal{V} \mathcal{P}}(Q^2) = (e_f - e_{\bar{f}^\prime})  I(q_\perp^2; M_\mathcal{P}, m_f) \;,
\label{TFF-equal-masses}
\end{equation}
where
\begin{equation}
	I (q_\perp^2; M_{\mathcal{P}},m_f) = \tilde{\mathcal{N}} \int \frac{\mathrm{d} x}{x} \left( \frac{M_{\mathcal{P}}}{2} + B \left(\frac{m_f}{x \bar{x}}\right) \right)\exp{\left(-\frac{(q_\perp \bar{x})^2 + 4 m_f^2}{4 \kappa^2 x\bar{x}}\right)} \;.
\label{integral-equal-mf}
\end{equation}
Eq. \eqref{TFF-equal-masses} is consistent with the model-independent expectation expressed by Eq. \eqref{minus-relation-Feynman-graphs} and it implies that $F_{\rho^\pm \pi^\pm}(Q^2)= F_{\rho^0 \pi^0}(Q^2)$ and $F_{\omega^0 \pi^0}(Q^2)= 3F_{\rho^0 \pi^0}(Q^2)$. For the strange mesons, we must instead use Eq. \eqref{TFF-int}, leading to the interesting possibility of destructive interference between the two Feynman diagrams of Fig. \ref{Fig:Feynman-diagrams} for the $K^{*\pm} \to K^\pm \gamma^*$ transition. We shall discuss this further in Section \ref{Sec:predictions}.

For the neutral mesons, $(\eta, \eta^\prime)$ and $(\phi,\omega)$, we need to account for  mixing. Although the $\phi-\omega$ mixing is small, it is essential to account for the $\phi \to \pi^0 \gamma^*$ transition. We use the SU(3) octet-singlet mixing scheme where 
\begin{eqnarray}
\left(\begin{array}{c}
         |\eta \rangle \\
         |\eta^\prime \rangle
    \end{array}\right)
&=&\left(\begin{array}{cc}
     \cos \theta_{\mathcal{P}} & -\sin \theta_{\mathcal{P}} \\
     \sin \theta_{\mathcal{P}} & \cos \theta_{\mathcal{P}} \\
   \end{array}\right)
    \left(\begin{array}{c}
          |\eta_8 \rangle \\
         |\eta_1 \rangle \\
     \end{array}\right),
     \end{eqnarray}
     \begin{eqnarray}          
\left(\begin{array}{c}
         |\phi \rangle \\
         |\omega \rangle
    \end{array}\right)
&=&\left(\begin{array}{cc}
     \cos \theta_{\mathcal{V}} & -\sin \theta_{\mathcal{V}} \\
     \sin \theta_{\mathcal{V}} & \cos \theta_{\mathcal{V}} \\
   \end{array}\right)
\left(\begin{array}{c}
     |\omega_8 \rangle \\
     |\omega_1 \rangle \\
   \end{array}\right) \;,
\end{eqnarray}
with $|\eta_8/\omega_8 \rangle=\frac{1}{\sqrt{6}}(u\bar{u}+ d\bar{d} -2 s\bar{s})$ and 
$|\eta_1/\omega_1 \rangle=\frac{1}{\sqrt{3}}(u\bar{u}+ d\bar{d} + s\bar{s})$. It then follows that \cite{Qian:2008px}
\begin{eqnarray}
\left(\begin{array}{c}
         F_{\rho^0\eta}(Q^2) \\
         F_{\eta\prime \rho^0}(Q^2)
    \end{array}\right)
&=&\left(\begin{array}{cc}
     \cos \theta_{\mathcal{P}} & -\sin \theta_{\mathcal{P}} \\
     \sin \theta_{\mathcal{P}} & \cos \theta_{\mathcal{P}} \\
   \end{array}\right)
    \left(\begin{array}{c}
          F_{\rho^0  \eta_8 }(Q^2)\\
          F_{\rho^0 \eta_1 }(Q^2) \\
     \end{array}\right),
     \end{eqnarray}
     \begin{eqnarray}          
\left(\begin{array}{c}
         F_{\phi \pi^0 }(Q^2) \\
         F_{\omega \pi^0}(Q^2)
    \end{array}\right)
&=&\left(\begin{array}{cc}
     \cos \theta_{\mathcal{V}} & -\sin \theta_{\mathcal{V}} \\
     \sin \theta_{\mathcal{V}} & \cos \theta_{\mathcal{V}} \\
   \end{array}\right)
\left(\begin{array}{c}
     F_{\omega_8 \pi^0}(Q^2)\\
     F_{\omega_1 \pi^0}(Q^2)\\
   \end{array}\right) \;,
\end{eqnarray}
and
\begin{eqnarray}
\nonumber
\left(
  \begin{array}{c}
    F_{\phi\eta} (Q^2)\\
    F_{\phi\eta\prime} (Q^2)\\
    F_{\omega\eta}(Q^2)\\
    F_{\eta\prime\omega}(Q^2)
  \end{array}
\right) &=& \left(
  \begin{array}{cccc}
   \cos\theta_{\mathcal{V}}\cos\theta_\mathcal{P} & -\cos\theta_{\mathcal{V}}\sin\theta_\mathcal{P} & -\sin\theta_{\mathcal{V}}\cos\theta_\mathcal{P} & \sin\theta_{\mathcal{V}}\sin\theta_\mathcal{P}\\
    \cos\theta_{\mathcal{V}}\sin\theta_\mathcal{P} & \cos\theta_{\mathcal{V}}\cos\theta_\mathcal{P} & -\sin\theta_{\mathcal{V}}\sin\theta_\mathcal{P} & -\sin\theta_{\mathcal{V}}\cos\theta_\mathcal{P}
\\
    \sin\theta_{\mathcal{V}}\cos\theta_\mathcal{P} & -\sin\theta_{\mathcal{V}}\sin\theta_\mathcal{P} & \cos\theta_{\mathcal{V}}\cos\theta_\mathcal{P} & -\cos\theta_{\mathcal{V}}\sin\theta_\mathcal{P}
\\
  \sin\theta_{\mathcal{V}}\sin\theta_\mathcal{P} & \sin\theta_{\mathcal{V}}\cos\theta_\mathcal{P} & \cos\theta_{\mathcal{V}}\sin\theta_\mathcal{P} & \cos\theta_{\mathcal{V}}\cos\theta_\mathcal{P}    
  \end{array}
\right)\left(
  \begin{array}{c}
    F_{\omega_8\eta_8}(Q^2) \\
    F_{\omega_8\eta_1}(Q^2) \\
    F_{\omega_1\eta_8}(Q^2) \\
    F_{\omega_1\eta_1}(Q^2) \\
  \end{array}
\right),
\end{eqnarray}
where, using Eq. \eqref{TFF-equal-masses},
\begin{eqnarray}
F_{\rho^0\eta_8}(Q^2) &=& \frac{1}{\sqrt{3}}I(q_\perp^2; M_{\eta_8}, m_q)\\
F_{\rho^0 \eta_1}(Q^2) &=& \sqrt{\frac{2}{3}}I(q_\perp^2; M_{\eta_1}, m_q)
   \\
F_{\omega_8\pi^0}(Q^2) &=&
    \frac{1}{\sqrt{3}}~I(q_\perp^2; M_\pi, m_q)\\
F_{\omega_1\pi^0}(Q^2) &=&
   \sqrt{\frac{2}{3}}I(q_\perp^2, M_\pi, m_q)\\
F_{\omega_8\eta_8}(Q^2) &=&
    \frac{1}{9}~I(q_\perp^2; M_{\eta_8},m_q)
    -\frac{4}{9}~I(q_\perp^2; M_{\eta_8},m_s)\\
F_{\omega_8\eta_1}(Q^2) &=&
    \frac{\sqrt{2}}{9}~I(q_\perp^2; M_{\eta_1},m_q)
    +\frac{2 \sqrt{2}}{9}~I(q_\perp^2; M_{\eta_1},m_s)\\
F_{\omega_1\eta_8}(Q^2) &=&
  \frac{\sqrt{2}}{9}~I(q_\perp^2; M_{\eta_8},m_q)
    +\frac{2 \sqrt{2}}{9}~I(q_\perp^2; M_{\eta_8},m_s)\\
F_{\omega_1\eta_1}(Q^2) &=&
    \frac{2}{9}~I(q_\perp^2; M_{\eta_1},m_q)
    -\frac{2}{9}~I(q_\perp^2; M_{\eta_1},m_s) \;,
\end{eqnarray}
with \cite{Ahmady:2018muv}
\begin{eqnarray}\left(
 \begin{array}{c}
   M^2_{\eta_8} \\
   M^2_{\eta_1}
  \end{array}
\right) &=& \left(
\begin{array}{cc}
\cos^2 \theta_\mathcal{P} & \sin^2 \theta_\mathcal{P}\\
\sin^2 \theta_\mathcal{P} & \cos^2 \theta_\mathcal{P}
\end{array}
\right) \left(
 \begin{array}{c}
   M^2_{\eta} \\
   M^2_{\eta\prime}
  \end{array}
\right) \;.
\end{eqnarray}
For a detailed analysis of mixing in the pseudoscalar sector, we refer to \cite{Feldmann:1998vh}.

Evaluating the transition form factors at $Q^2=0$ allow us to predict the radiative decay widths:
\begin{equation}
	\Gamma_{\mathcal{V} \to \mathcal{P} \gamma} = \frac{\alpha_{\mathrm{em}}}{3} |F_{\mathcal{V} \mathcal{P}}(0)|^2 \left( \frac{M_{\mathcal{V}}^2-M_{\mathcal{P}}^2}{2M_\mathcal{V}} \right)^3 \;,
\label{decay-widths}
\end{equation} 
and, as mentioned before, to predict the timelike transition form factor, we use the prescription $q_\perp \to i q_\perp$ in Eq. \eqref{integral} which then reads: 
\begin{equation}
	I (q_\perp^2; M_{\mathcal{P}}, m_f, m_{\bar{f}^\prime}) = \int  \frac{\mathrm{d} x}{x} \left( \frac{M_{\mathcal{P}}}{2} + B \left(\frac{\bar{x} m_f+ x m_{\bar{f}}}{x\bar{x}}\right) \right)	 \exp{\left(\frac{(q_\perp \bar{x})^2 - 4 (\bar{x} m_f^2 + x m^2_{\bar{f}})}{4 \kappa^2 x\bar{x}}\right)} \;.
\label{integral-timelike}
\end{equation}
As expected, Eq. \eqref{integral-timelike} diverges for $q_\perp^2 \ge 4m^2_{u/d}$, corresponding to the kinematic threshold for quark-antiquark production. Since we do not account for the latter here, we shall restrict our predictions in the timelike region below this threshold. 

In order to reproduce the non-perturbative pole structure of the form factor in the timelike region, above the quark-antiquark production threshold, one must use the confined bulk-to-boundary propagator, i.e. one which propagates in the dilation-modified $\mathrm{AdS}_5$ spacetime and maps onto a ``dressed" (i.e. incorporating higher Fock states) electromagnetic current in physical spacetime \cite{Brodsky:2014yha}. The resulting form factor also reproduces the VMD behaviour in the low momentum region, as well as the hard scattering power scaling behaviour at large $Q^2$. This technique has been used to predict the pion electromagnetic form factor \cite{Brodsky:2007hb,Brodsky:2014yha}, the $(\pi^0, \eta, \eta^\prime) \to \gamma^* \gamma$ transition form factors \cite{Brodsky:2011xx,Brodsky:2011yv} as well as the nucleon electromagnetic form factors in the spacelike region \cite{Chakrabarti:2013dda,Sufian:2016hwn}.

\section{Comparing to data}
\label{Sec:predictions}

For our numerical predictions, we use $m_{u/d}=330 \pm 30$ MeV, $m_s=500 \pm 30$ MeV and the universal AdS/QCD scale, $\kappa=523 \pm 24$ MeV,  as in Ref. \cite{Ahmady:2018muv}. For the mixing angles, we use $\theta_{\mathcal{P}}=-(14.1 \pm 2.8)^\circ$\cite{Christ:2010dd} and $\theta_{\mathcal{V}}=(38.7 \pm 0.2)^\circ$ \cite{Escribano:2005qq}. Our theory uncertainties follow from these quoted uncertainties.

The various experimental collaborations fit the timelike transition form factor data using
\begin{equation}
	|F_{\mathrm{exp}}(Q^2)|^2=\frac{1}{\left(1+\frac{Q^2}{\Lambda^2}\right)^2}
\label{Fexp}
\end{equation}
where $\Lambda$ is the parameter to be fitted. Reported values are: $\Lambda_{\mathrm{NA60}}=0.670 \pm 0.006$ GeV, $\Lambda_{\mathrm{A2MM}}=0.709 \pm 0.037$ GeV and  $\Lambda_{\mathrm{Lepton-G}}=0.65 \pm 0.037$ GeV for the $\omega \to \pi^0 \gamma^*$ transition, $\Lambda_{\mathrm{KLOE}}=0.704 \pm 0.019$ for the $\phi \to \pi^0 \gamma^*$ transition, and $\Lambda_{\mathrm{PDG}}=0.88 \pm 0.04$ GeV for the $\phi \to \eta \gamma^*$ transition.  Note that, with $\Lambda=M_\rho$, Eq. \eqref{Fexp} is the VMD prediction.

\begin{figure}[hbt]
\centering
\includegraphics[width=15cm]{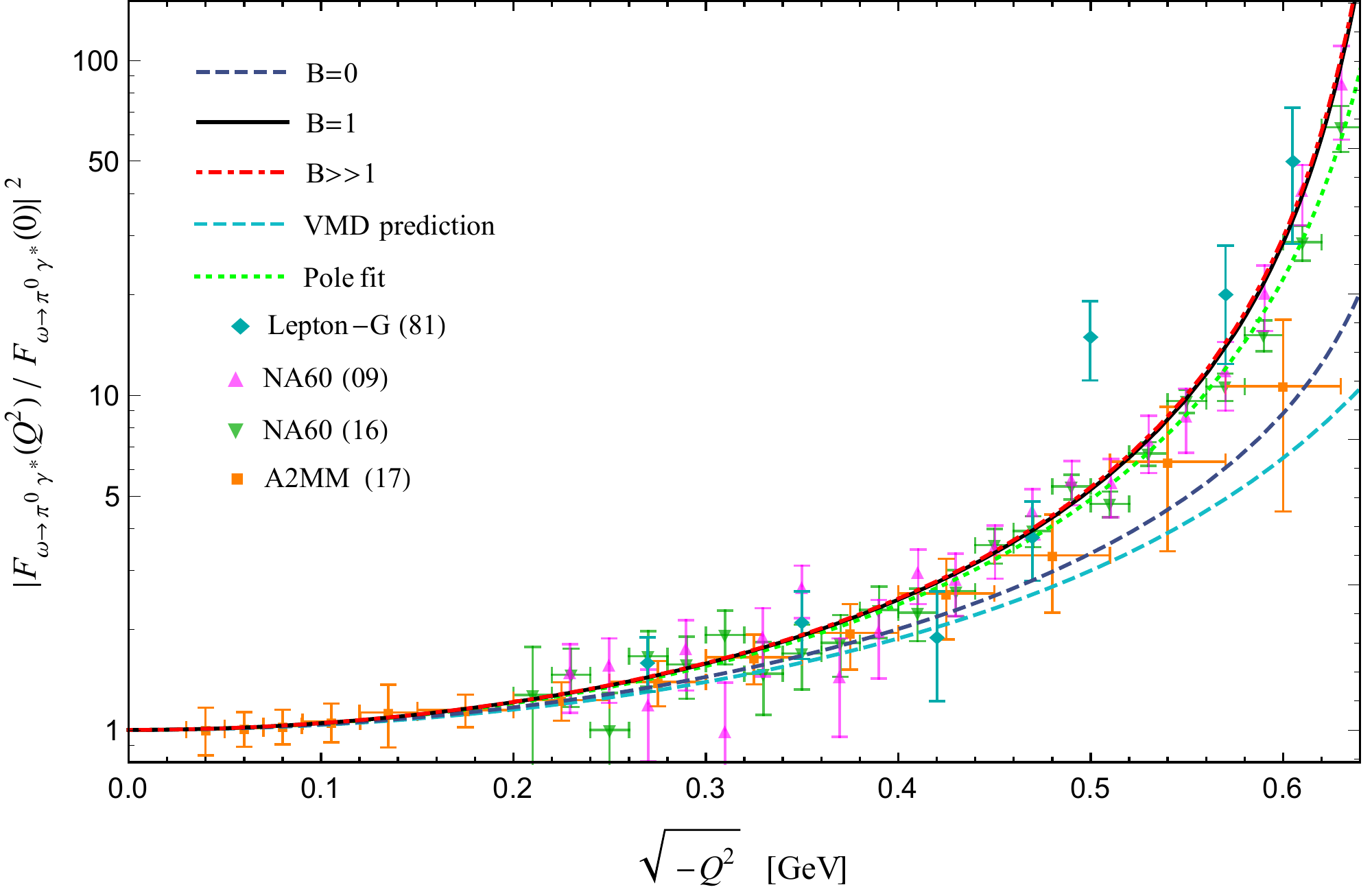}
\caption{Our predictions for the $\omega \to \pi^0 \gamma^*$ timelike transition form factor with $B=0$ (dashed-blue curve), $B = 1$ (solid-black curve) and $B \gg 1$ (dot-dashed red curve), compared to the data from Refs. \cite{Dzhelyadin:1980tj,Adlarson:2016hpp,Arnaldi:2016pzu,Arnaldi:2009aa}. The dashed-cyan curve is the VMD prediction and the empirical pole fit (dotted-green curve) is generated by Eq. \eqref{Fexp}, with $\Lambda$ being the averages the fitted values reported by Lepton-G, A2MM and NA60 experiments.}
   \label{Fig:omega}  
   \end{figure}

\begin{figure}[hbt]
 \includegraphics[width=15cm]{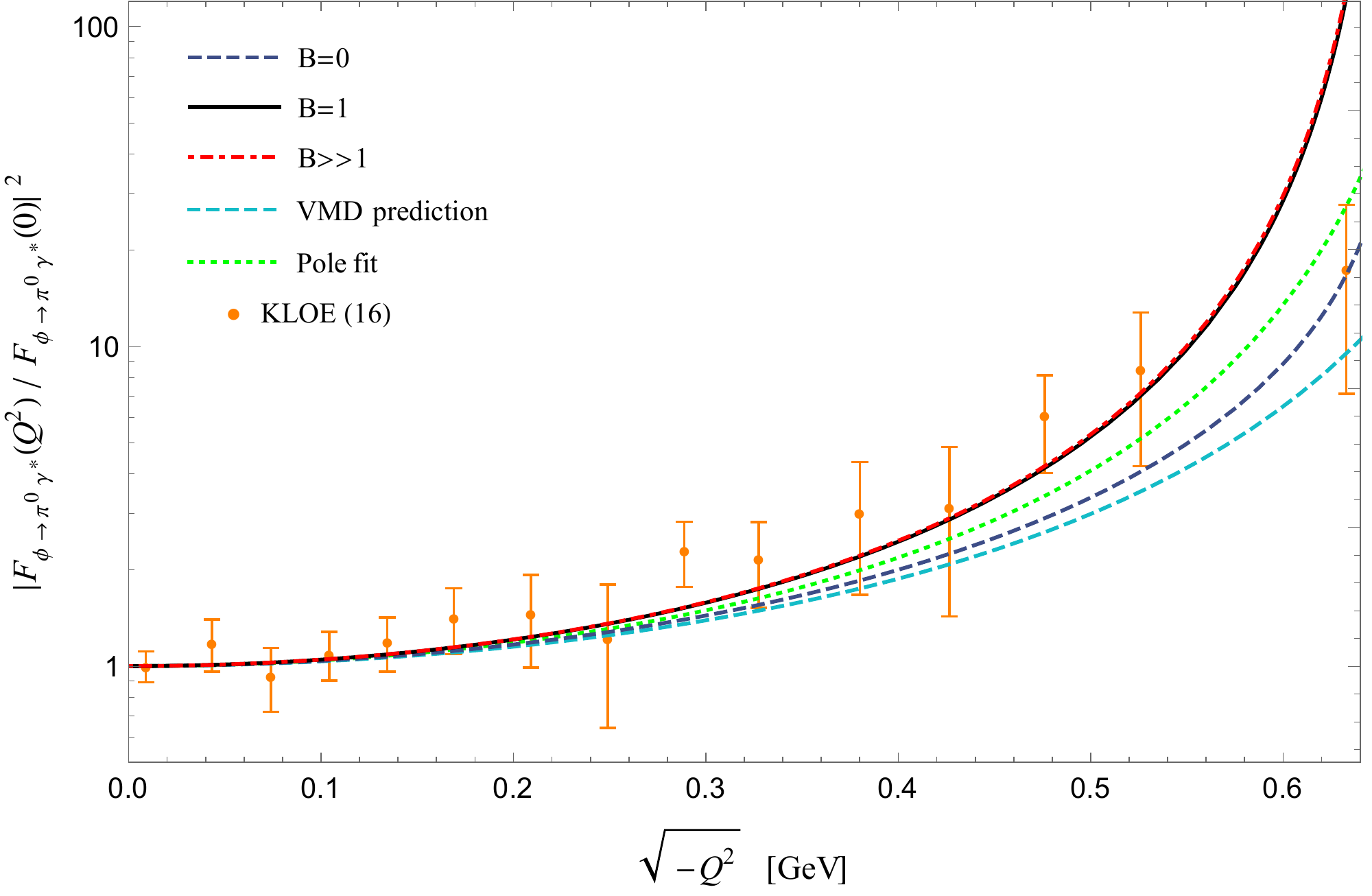}
   \caption{Our predictions for the $\phi \to \pi^0 \gamma^*$ timelike transition form factor, with $B=0$ (dashed-blue curve), $B = 1$ (solid-black curve) and $B \gg 1$ (dot-dashed-red curve), compared to the data from Ref. \cite{Anastasi:2016qga}. The dashed-cyan curve is the VMD prediction and the empirical pole fit (dotted-green curve) is generated by Eq. \eqref{Fexp} with  $\Lambda=\Lambda_{\mathrm{KLOE}}$.}
   \label{Fig:phi}  
   \end{figure}

Our predictions for the $(\rho, \omega, \phi) \to \pi \gamma$ radiative decay widths are shown in Table \ref{Tab:pions}. As can be seen, $B \ge 1$ is favoured by the data, corroborating the findings of Ref. \cite{Ahmady:2018muv} that $B \ge 1$ is favoured for the pion. This is further supported by our predictions for the $\omega \to \pi^0 \gamma^*$ timelike transition form factor, as shown in Fig. \ref{Fig:omega}. The empirical pole fit (dotted-green curve) is generated using Eq. \eqref{Fexp} with $\Lambda=0.676$ GeV, the average of the Lepton-G, A2MM and NA60 values, and it agrees very well with our $B\ge 1$ predictions (solid-black and dot-dashed red curves). Our predictions for the $\phi \to \pi^0 \gamma^*$ timelike transition form factor are shown in Fig. \ref{Fig:phi}. In this case, although there is a preference for the $B \ge 1$ predictions (solid-black and dot-dashed red curves), the larger error bars of the data do not completely exclude the $B=0$ (dashed-blue) prediction. Indeed, the empirical pole fit (dotted-green curve) now lies between the $B \ge 1$ (solid-black and dot-dashed-red) and $B=0$ (dashed-blue) curves. The predictions with $B \ge 1$ are particularly impressive since they can be viewed as parameter-free: once $B \ge 1$ is fixed, as in Ref. \cite{Ahmady:2018muv}, all other predictions are obtained without any further adjustment of parameters.  

 \begin{table}
 \begin{tabular}{|  c  |  c  |  c  |  c  | c  |}
 \cline{1-5}
 & \multicolumn{3}{c|}{Spin-improved LFH [keV]} &\\ \cline{2-4}
~~~~~~~ Decay widths ~~~~~~~&~~~~~~~~ ${\rm B}=0$~~~~~~~~ & ~~~~~~~~${\rm B}=1$~~~~~~~~ & ~~~~~~~${\rm B}>>1$ ~~~~~~~ & PDG (2018) [keV] \\ \hline \hline 
 $\Gamma(\rho^\pm \rightarrow \pi^\pm \gamma)$ & $ 23.46 \pm 3.12$ & $64.52 \pm 6.94$ & $66.37 \pm 7.00$ & $67.10 \pm 7.82$  \\ \cline{1-5}
 
$\Gamma(\rho^0 \rightarrow \pi^0 \gamma)$ & $23.46 \pm 3.12$ & $64.52 \pm 6.94$ & $66.37 \pm 7.00$ & $70.08 \pm 9.32$  \\ \cline{1-5}

 $\Gamma(\omega \rightarrow \pi^0 \gamma)$ & $221.03 \pm 29.90$ & $607.96 \pm 65.44$ & $625.38 \pm 66.03$ & $713.16 \pm 25.40$ \\ \cline{1-5}
 
  $\Gamma(\phi \rightarrow \pi^0 \gamma)$ &$1.84 \pm 0.33$ & $5.06 \pm 0.80$ & $5.21 \pm 0.82$ & $5.52 \pm 0.22$ \\ \cline{1-5}
  
  \end{tabular}
  \caption{Our predictions for the $(\rho,\omega,\phi) \to \pi \gamma$ decay widths, compared to the PDG averages \cite{PDG}.}
  \label{Tab:pions}
  \end{table}

\begin{figure}[hbt]
 \includegraphics[width=15cm]{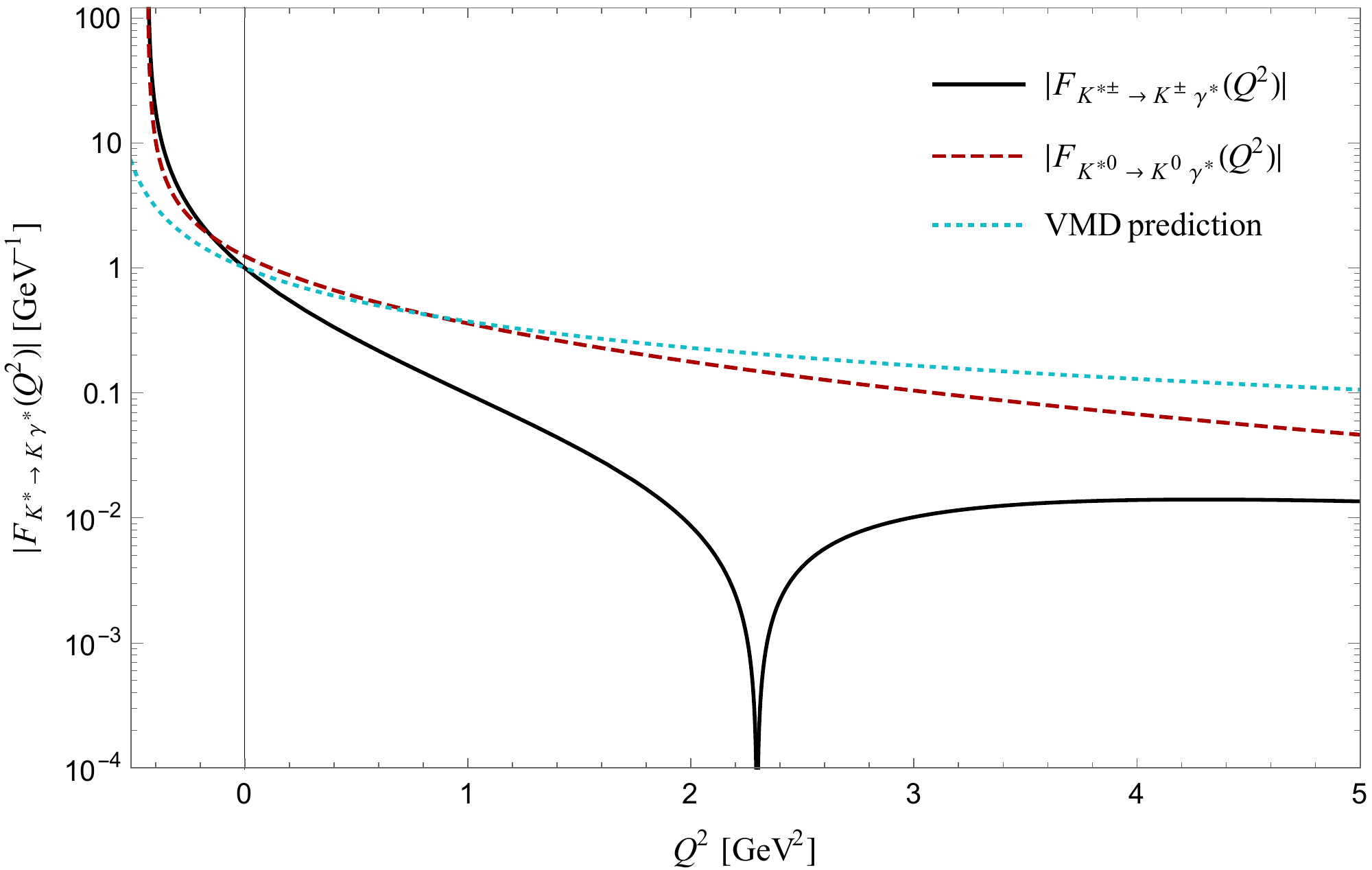}
   \caption{Our predictions for the $K^{*\pm} \to K^\pm \gamma^*$ (solid-black curve) and $K^{*0} \to K^0 \gamma^*$ (dashed-brown curve) transition form factors, with $B=1$, compared to the VMD prediction (dotted-cyan curve).}
   \label{Fig:kaons}  
   \end{figure}

For the $K^{*0} \to K^0 \gamma$ and $K^{*\pm} \to K^{\pm} \gamma$ decay widths, Table \ref{Tab:kaons} shows that $B=1$ accommodates the data for both the neutral and charged decay modes. Note that the theory uncertainty is amplified for the latter because of the destructive interference between the two Feynman graphs of Fig. \ref{Fig:Feynman-diagrams}. At first glance, the preference for $B=1$ for the charged decay mode may seem in disagreement with the findings of Ref. \cite{Ahmady:2018muv}, where $B=0$ is reported to be preferred by decay constant, electromagnetic elastic form factor and radius data for charged kaons. However, we must emphasize that taking $0 < B \ll 1$, say $B=0.2$, still fits the radiative width data in Table \ref{Tab:kaons}, as well as all data in Ref. \cite{Ahmady:2018muv}.  On the other hand, as can be seen in Table \ref{Tab:kaons}, $B < 1$ is excluded for the neutral decay mode. As we mentioned before, destructive interference occurs only in the charged decay mode, leading to a zero (at leading order) in the transition form factor in the spacelike region. This is shown in Fig. \ref{Fig:kaons}.  We note that the location of the zero is sensitive to the strength of SU(3) flavour symmetry breaking, shifting to lower $Q^2$ as the difference between $m_s$ and $m_q$ increases, as was pointed out previously in Refs. \cite{Munz:1994ty,Choi:2008si}, although the precise location of the zero is very much model-dependent.

 \begin{table}
 \begin{tabular}{|  c  |  c  |  c  |  c  | c  |}
 \cline{1-5}
 & \multicolumn{3}{c|}{Spin-improved LFH [keV]} &\\ \cline{2-4}
~~~~~~~ Decay widths ~~~~~~~&~~~~~~~~ ${\rm B}=0$~~~~~~~~ & ~~~~~~~~${\rm B}=1$~~~~~~~~ & ~~~~~~~${\rm B}>>1$ ~~~~~~~ & PDG (2018) [keV] \\ \hline \hline 
 $\Gamma(K^{*0} \rightarrow K^0 \gamma)$ & $39.38 \pm 3.74$ & $108.67 \pm 9.34$ & $122.02 \pm 10.49$ & $116.36 \pm 11.17$ \\ \cline{1-5} 
 
  $\Gamma(K^{* \pm} \rightarrow K^{\pm} \gamma)$ & $23.85 \pm 5.74$ & $71.64 \pm 18.17$ & $81.20 \pm 20.66 $ & $50.29 \pm 5.47$ \\ \cline{1-5}
\end{tabular}
\caption{Our predictions for the $K^* \to K \gamma$ decay widths, compared to the PDG averages \cite{PDG}.}
  \label{Tab:kaons}
  \end{table}

\begin{figure}[hbt]
 \includegraphics[width=15cm]{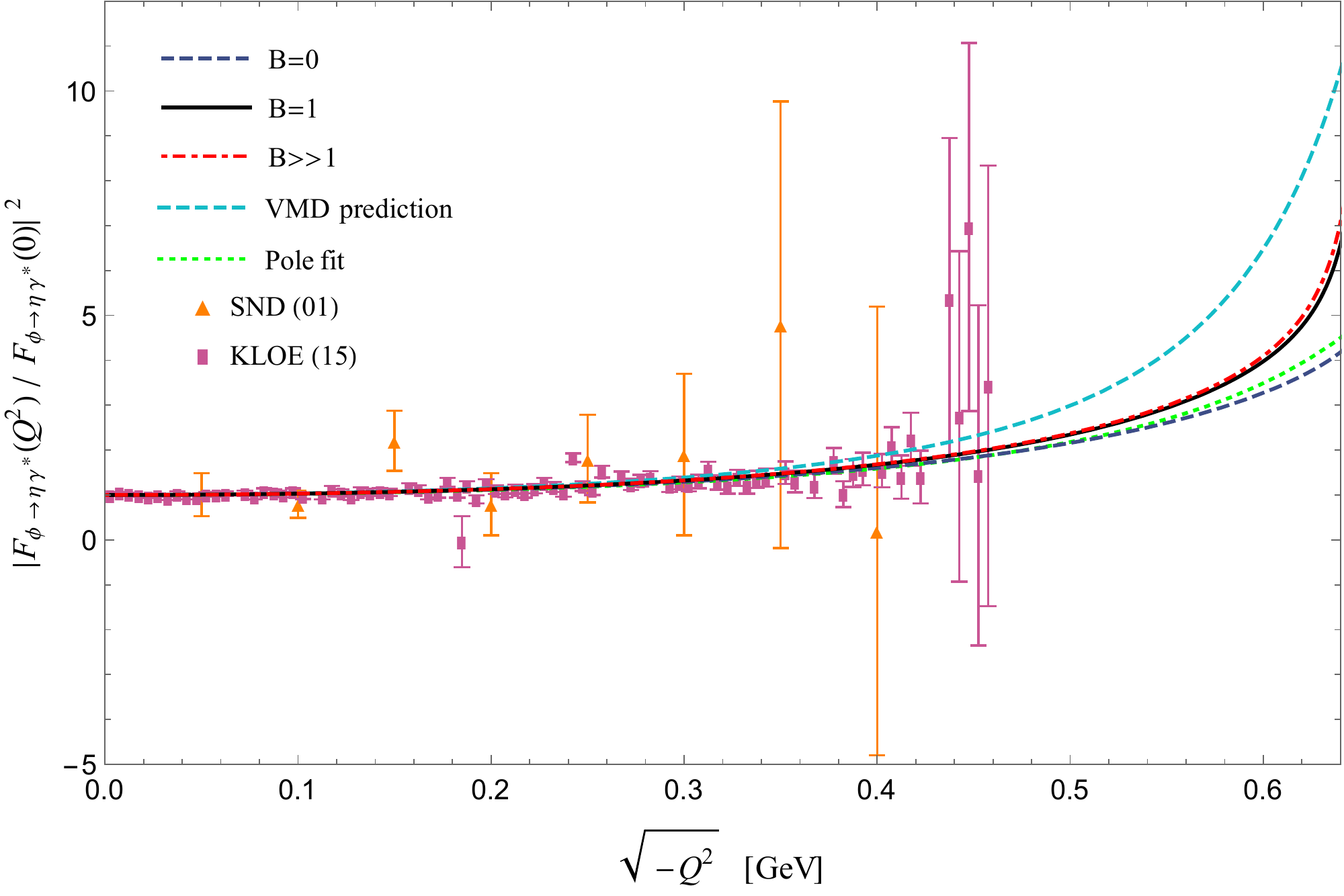}
   \caption{Our predictions for the $\phi \to \eta \gamma^*$ timelike transition form factor with $B=0$ (dashed-blue curve), $B = 1$ (solid-black curve) and $B \gg 1$ (dot-dashed-red curve), compared to the data from Refs. \cite{Achasov:2000ne, Babusci:2014ldz}. The dashed-cyan curve is the VMD prediction and the empirical pole fit (dotted-green curve) is generated by Eq. \eqref{Fexp},  with $\Lambda=\Lambda_{\mathrm{PDG}}$.}
   \label{Fig:eta}  
   \end{figure}

In Table \ref{Tab:eta}, we show our predictions for the radiative decays to $\eta$ and $\eta^\prime$ where an additional theory uncertainty results from the $\eta/\eta^\prime$ mixing angle. Clearly, $B \ge 1$ is preferred by the data. This is consistent with the findings of Ref. \cite{Ahmady:2018muv} where it is reported that $B \ge 1$ is also preferred by the $\eta/\eta^\prime \to \gamma^* \gamma$ transition form factor data.  In Fig. \ref{Fig:eta}, we compare our predictions for the $\phi \to \eta \gamma^*$ transition to KLOE and SND data. In this case, the data cannot discriminate between the $B=0$ (dashed-blue curve) and $B \ge 1$ (solid-black and dot-dashed-red curves) predictions which start to differ only at large momentum transfer where the experimental error bars are much larger. Both the $B=0$ and $B \ge 1$ curves agree with the empirical pole fit (dotted-green curve) which is now generated with $\Lambda_{\mathrm{PDG}}=0.88 \pm 0.04$ GeV. Finally, we also predict the $\eta^\prime \to (\rho,\omega) \gamma$ decay widths given by \begin{equation}
 \Gamma_{\eta^\prime \to (\rho,\omega) \gamma} = \alpha_{\mathrm{em}} |F_{\mathcal{\eta^\prime} \mathcal{(\rho,\omega)}}(0)|^2 \left( \frac{M_{\mathcal{\eta^\prime}}^2-M_{\mathcal{(\rho,\omega)}}^2}{2M_\mathcal{\eta^\prime}} \right)^3 \;.
 \end{equation}
 Our results are shown in Table \ref{Tab:eta-prime} where we find that $B \ge 1$ is again favoured by the data.

 \begin{table}
 \begin{tabular}{|  c  |  c  |  c  |  c  | c  |}
 \cline{1-5}
 & \multicolumn{3}{c|}{Spin-improved LFH [keV]} &\\ \cline{2-4}
~~~~~~~ Decay widths ~~~~~~~&~~~~~~~~ ${\rm B}=0$~~~~~~~~ & ~~~~~~~~${\rm B}=1$~~~~~~~~ & ~~~~~~~${\rm B}>>1$ ~~~~~~~ & PDG (2018) [keV] \\ \hline \hline 
  $\Gamma(\rho \rightarrow \eta \gamma)$ & $16.18 \pm 2.57$ & $40.00 \pm 5.50$ & $45.73 \pm 6.16$ & $44.70 \pm 3.37$  \\ \cline{1-5}
  
  $\Gamma(\omega \rightarrow \eta \gamma)$ & $1.76 \pm 0.31 $ & $4.31 \pm 0.67$ & $4.93 \pm 0.75$ & $3.82 \pm 0.38$ \\ \cline{1-5}
  
   $\Gamma(\phi \rightarrow \eta \gamma)$ & $20.80 \pm 3.01$ & $59.64 \pm 8.07$  & $67.63 \pm 9.21$ & $55.36 \pm 1.23$ \\ \cline{1-5}

  $\Gamma(\phi \rightarrow \eta \prime \gamma)$ & $0.11 \pm 0.02$ & $0.29 \pm 0.04$ & $0.36 \pm 0.05$ & $0.26 \pm 0.01$ \\ \cline{1-5}
  
  \end{tabular}
  \caption{Our predictions for the $(\rho,\omega,\phi) \to (\eta,\eta^\prime) \gamma$ decay widths, compared to the PDG averages \cite{PDG}.}
  \label{Tab:eta}  
\end{table}

 \begin{table}
 \begin{tabular}{|  c  |  c  |  c  |  c  | c  |}
 \cline{1-5}
 & \multicolumn{3}{c|}{Spin-improved LFH [keV]} &\\ \cline{2-4}
~~~~~~~ Decay widths ~~~~~~~&~~~~~~~~ ${\rm B}=0$~~~~~~~~ & ~~~~~~~~${\rm B}=1$~~~~~~~~ & ~~~~~~~${\rm B}>>1$ ~~~~~~~ & PDG (2018) [keV] \\ \hline \hline 
  $\Gamma(\eta \prime \rightarrow \rho \gamma)$ & $25.38 \pm 4.48$ & $58.80 \pm 9.54$ & $71.68 \pm 11.13$ &  $56.64 \pm 3.58$ \\ \cline{1-5}

   $\Gamma(\eta \prime \rightarrow \omega \gamma)$ & $2.85 \pm 0.45$ & $6.70 \pm 0.95$ & $8.16 \pm 1.11$ & $5.14 \pm 0.49$ \\ \cline{1-5}
\end{tabular}
\caption{Our predictions for the $\eta^\prime \to (\rho,\omega) \gamma$ decay widths, compared to the PDG averages \cite{PDG}.}
  \label{Tab:eta-prime}  
\end{table}

 \section{Conclusions}
 We have used the spin-improved holographic light-front wavefunctions for the light vector mesons $(\rho, \omega, K^*, \phi)$ and pseudoscalar mesons $(\pi, K, \eta,\eta^\prime)$ to predict the radiative transition form factors and decay widths. We find excellent agreement with the available data for the decay widths as well as the timelike transition form factors in the low-momentum region. Our findings support the idea that light pseudoscalar and vector mesons share a universal holographic light-front wavefunction which is modified differently by dynamical spin effects.

   \section{Acknowledgements}
M.A and R.S are supported by Individual Discovery Grants from the Natural Science and Engineering Research Council of Canada (NSERC): SAPIN-2017-00033 and SAPIN-2020-00051 respectively. C.M is supported by the Natural Science Foundation of China (NSFC) under the grants No. 11850410436 and 11950410753. S.K is supported by an Institute Fellowship from the Ministry of Human Resource Development (MHRD), Government of India.
\bibliographystyle{apsrev}
\bibliography{TFF-revised}

\begin{thebibliography}{63}
\expandafter\ifx\csname natexlab\endcsname\relax\def\natexlab#1{#1}\fi
\expandafter\ifx\csname bibnamefont\endcsname\relax
  \def\bibnamefont#1{#1}\fi
\expandafter\ifx\csname bibfnamefont\endcsname\relax
  \def\bibfnamefont#1{#1}\fi
\expandafter\ifx\csname citenamefont\endcsname\relax
  \def\citenamefont#1{#1}\fi
\expandafter\ifx\csname url\endcsname\relax
  \def\url#1{\texttt{#1}}\fi
\expandafter\ifx\csname urlprefix\endcsname\relax\def\urlprefix{URL }\fi
\providecommand{\bibinfo}[2]{#2}
\providecommand{\eprint}[2][]{\url{#2}}

\bibitem[{\citenamefont{Jegerlehner and Nyffeler}(2009)}]{Jegerlehner:2009ry}
\bibinfo{author}{\bibfnamefont{F.}~\bibnamefont{Jegerlehner}} \bibnamefont{and}
  \bibinfo{author}{\bibfnamefont{A.}~\bibnamefont{Nyffeler}},
  \bibinfo{journal}{Phys. Rept.} \textbf{\bibinfo{volume}{477}},
  \bibinfo{pages}{1} (\bibinfo{year}{2009}), \eprint{0902.3360}.

\bibitem[{\citenamefont{Tanabashi et~al.}(2018)\citenamefont{Tanabashi,
  Hagiwara, Hikasa, Nakamura, Sumino, Takahashi, Tanaka, Agashe, Aielli, Amsler
  et~al.}}]{PDG}
\bibinfo{author}{\bibfnamefont{M.}~\bibnamefont{Tanabashi}},
  \bibinfo{author}{\bibfnamefont{K.}~\bibnamefont{Hagiwara}},
  \bibinfo{author}{\bibfnamefont{K.}~\bibnamefont{Hikasa}},
  \bibinfo{author}{\bibfnamefont{K.}~\bibnamefont{Nakamura}},
  \bibinfo{author}{\bibfnamefont{Y.}~\bibnamefont{Sumino}},
  \bibinfo{author}{\bibfnamefont{F.}~\bibnamefont{Takahashi}},
  \bibinfo{author}{\bibfnamefont{J.}~\bibnamefont{Tanaka}},
  \bibinfo{author}{\bibfnamefont{K.}~\bibnamefont{Agashe}},
  \bibinfo{author}{\bibfnamefont{G.}~\bibnamefont{Aielli}},
  \bibinfo{author}{\bibfnamefont{C.}~\bibnamefont{Amsler}},
  \bibnamefont{et~al.} (\bibinfo{collaboration}{Particle Data Group}),
  \bibinfo{journal}{Phys. Rev. D} \textbf{\bibinfo{volume}{98}},
  \bibinfo{pages}{030001} (\bibinfo{year}{2018}).

\bibitem[{\citenamefont{Dzhelyadin et~al.}(1981)}]{Dzhelyadin:1980tj}
\bibinfo{author}{\bibfnamefont{R.}~\bibnamefont{Dzhelyadin}}
  \bibnamefont{et~al.}, \bibinfo{journal}{JETP Lett.}
  \textbf{\bibinfo{volume}{33}}, \bibinfo{pages}{228} (\bibinfo{year}{1981}).

\bibitem[{\citenamefont{Arnaldi et~al.}(2009)}]{Arnaldi:2009aa}
\bibinfo{author}{\bibfnamefont{R.}~\bibnamefont{Arnaldi}} \bibnamefont{et~al.}
  (\bibinfo{collaboration}{NA60}), \bibinfo{journal}{Phys. Lett.}
  \textbf{\bibinfo{volume}{B677}}, \bibinfo{pages}{260} (\bibinfo{year}{2009}),
  \eprint{0902.2547}.

\bibitem[{\citenamefont{Arnaldi et~al.}(2016)}]{Arnaldi:2016pzu}
\bibinfo{author}{\bibfnamefont{R.}~\bibnamefont{Arnaldi}} \bibnamefont{et~al.}
  (\bibinfo{collaboration}{NA60}), \bibinfo{journal}{Phys. Lett.}
  \textbf{\bibinfo{volume}{B757}}, \bibinfo{pages}{437} (\bibinfo{year}{2016}),
  \eprint{1608.07898}.

\bibitem[{\citenamefont{Adlarson et~al.}(2017)}]{Adlarson:2016hpp}
\bibinfo{author}{\bibfnamefont{P.}~\bibnamefont{Adlarson}}
  \bibnamefont{et~al.}, \bibinfo{journal}{Phys. Rev.}
  \textbf{\bibinfo{volume}{C95}}, \bibinfo{pages}{035208}
  (\bibinfo{year}{2017}), \eprint{1609.04503}.

\bibitem[{\citenamefont{Achasov et~al.}(2001)}]{Achasov:2000ne}
\bibinfo{author}{\bibfnamefont{M.~N.} \bibnamefont{Achasov}}
  \bibnamefont{et~al.}, \bibinfo{journal}{Phys. Lett.}
  \textbf{\bibinfo{volume}{B504}}, \bibinfo{pages}{275} (\bibinfo{year}{2001}).

\bibitem[{\citenamefont{Babusci et~al.}(2015)}]{Babusci:2014ldz}
\bibinfo{author}{\bibfnamefont{D.}~\bibnamefont{Babusci}} \bibnamefont{et~al.}
  (\bibinfo{collaboration}{KLOE-2}), \bibinfo{journal}{Phys. Lett.}
  \textbf{\bibinfo{volume}{B742}}, \bibinfo{pages}{1} (\bibinfo{year}{2015}),
  \eprint{1409.4582}.

\bibitem[{\citenamefont{Terschl{\"u}sen and
  Leupold}(2010)}]{TERSCHLUSEN2010191}
\bibinfo{author}{\bibfnamefont{C.}~\bibnamefont{Terschl{\"u}sen}}
  \bibnamefont{and} \bibinfo{author}{\bibfnamefont{S.}~\bibnamefont{Leupold}},
  \bibinfo{journal}{Physics Letters B} \textbf{\bibinfo{volume}{691}},
  \bibinfo{pages}{191 } (\bibinfo{year}{2010}), ISSN \bibinfo{issn}{0370-2693}.

\bibitem[{\citenamefont{Schneider et~al.}(2012)\citenamefont{Schneider, Kubis,
  and Niecknig}}]{Schneider:2012ez}
\bibinfo{author}{\bibfnamefont{S.~P.} \bibnamefont{Schneider}},
  \bibinfo{author}{\bibfnamefont{B.}~\bibnamefont{Kubis}}, \bibnamefont{and}
  \bibinfo{author}{\bibfnamefont{F.}~\bibnamefont{Niecknig}},
  \bibinfo{journal}{Phys. Rev. D} \textbf{\bibinfo{volume}{86}},
  \bibinfo{pages}{054013} (\bibinfo{year}{2012}), \eprint{1206.3098}.

\bibitem[{\citenamefont{Ivashyn}(2012)}]{Ivashyn:2011hb}
\bibinfo{author}{\bibfnamefont{S.}~\bibnamefont{Ivashyn}},
  \bibinfo{journal}{Prob. Atomic Sci. Technol.}
  \textbf{\bibinfo{volume}{2012N1}}, \bibinfo{pages}{179}
  (\bibinfo{year}{2012}), \eprint{1111.1291}.

\bibitem[{\citenamefont{Danilkin et~al.}(2015)\citenamefont{Danilkin,
  Fern\'andez-Ram\'{\i}rez, Guo, Mathieu, Schott, Shi, and
  Szczepaniak}}]{PhysRevD.91.094029}
\bibinfo{author}{\bibfnamefont{I.~V.} \bibnamefont{Danilkin}},
  \bibinfo{author}{\bibfnamefont{C.}~\bibnamefont{Fern\'andez-Ram\'{\i}rez}},
  \bibinfo{author}{\bibfnamefont{P.}~\bibnamefont{Guo}},
  \bibinfo{author}{\bibfnamefont{V.}~\bibnamefont{Mathieu}},
  \bibinfo{author}{\bibfnamefont{D.}~\bibnamefont{Schott}},
  \bibinfo{author}{\bibfnamefont{M.}~\bibnamefont{Shi}}, \bibnamefont{and}
  \bibinfo{author}{\bibfnamefont{A.~P.} \bibnamefont{Szczepaniak}},
  \bibinfo{journal}{Phys. Rev. D} \textbf{\bibinfo{volume}{91}},
  \bibinfo{pages}{094029} (\bibinfo{year}{2015}).

\bibitem[{\citenamefont{Anastasi et~al.}(2016)}]{Anastasi:2016qga}
\bibinfo{author}{\bibfnamefont{A.}~\bibnamefont{Anastasi}} \bibnamefont{et~al.}
  (\bibinfo{collaboration}{KLOE-2}), \bibinfo{journal}{Phys. Lett.}
  \textbf{\bibinfo{volume}{B757}}, \bibinfo{pages}{362} (\bibinfo{year}{2016}),
  \eprint{1601.06565}.

\bibitem[{\citenamefont{Maris and Tandy}(2002)}]{Maris:2002mz}
\bibinfo{author}{\bibfnamefont{P.}~\bibnamefont{Maris}} \bibnamefont{and}
  \bibinfo{author}{\bibfnamefont{P.~C.} \bibnamefont{Tandy}},
  \bibinfo{journal}{Phys. Rev. C} \textbf{\bibinfo{volume}{65}},
  \bibinfo{pages}{045211} (\bibinfo{year}{2002}), \eprint{nucl-th/0201017}.

\bibitem[{\citenamefont{Zhang et~al.}(2015)\citenamefont{Zhang, Cheng, Hua, and
  Xiao}}]{Zhang:2015mxa}
\bibinfo{author}{\bibfnamefont{Y.-L.} \bibnamefont{Zhang}},
  \bibinfo{author}{\bibfnamefont{S.}~\bibnamefont{Cheng}},
  \bibinfo{author}{\bibfnamefont{J.}~\bibnamefont{Hua}}, \bibnamefont{and}
  \bibinfo{author}{\bibfnamefont{Z.-J.} \bibnamefont{Xiao}},
  \bibinfo{journal}{Phys. Rev. D} \textbf{\bibinfo{volume}{92}},
  \bibinfo{pages}{094031} (\bibinfo{year}{2015}), \bibinfo{note}{[Addendum:
  Phys.Rev.D 93, 099901 (2016)]}, \eprint{1510.05108}.

\bibitem[{\citenamefont{Kimura et~al.}(2018)\citenamefont{Kimura, Morozumi, and
  Umeeda}}]{Kimura:2016xnx}
\bibinfo{author}{\bibfnamefont{D.}~\bibnamefont{Kimura}},
  \bibinfo{author}{\bibfnamefont{T.}~\bibnamefont{Morozumi}}, \bibnamefont{and}
  \bibinfo{author}{\bibfnamefont{H.}~\bibnamefont{Umeeda}},
  \bibinfo{journal}{PTEP} \textbf{\bibinfo{volume}{2018}},
  \bibinfo{pages}{123B02} (\bibinfo{year}{2018}), \eprint{1609.09235}.

\bibitem[{\citenamefont{Feldmann}(2000)}]{Feldmann:1999uf}
\bibinfo{author}{\bibfnamefont{T.}~\bibnamefont{Feldmann}},
  \bibinfo{journal}{Int. J. Mod. Phys.} \textbf{\bibinfo{volume}{A15}},
  \bibinfo{pages}{159} (\bibinfo{year}{2000}), \eprint{hep-ph/9907491}.

\bibitem[{\citenamefont{Choi and Ji}(1997)}]{Choi:1996mq}
\bibinfo{author}{\bibfnamefont{H.}~\bibnamefont{Choi}} \bibnamefont{and}
  \bibinfo{author}{\bibfnamefont{C.-R.} \bibnamefont{Ji}},
  \bibinfo{journal}{Nucl.\ Phys.\ A} \textbf{\bibinfo{volume}{618}},
  \bibinfo{pages}{291} (\bibinfo{year}{1997}).

\bibitem[{\citenamefont{Choi}(2008)}]{Choi:2008si}
\bibinfo{author}{\bibfnamefont{H.-M.} \bibnamefont{Choi}},
  \bibinfo{journal}{Phys. Rev. D} \textbf{\bibinfo{volume}{77}},
  \bibinfo{pages}{097301} (\bibinfo{year}{2008}), \eprint{0803.3100}.

\bibitem[{\citenamefont{Yu et~al.}(2007)\citenamefont{Yu, Xiao, and
  Ma}}]{Yu_2007}
\bibinfo{author}{\bibfnamefont{J.}~\bibnamefont{Yu}},
  \bibinfo{author}{\bibfnamefont{B.-W.} \bibnamefont{Xiao}}, \bibnamefont{and}
  \bibinfo{author}{\bibfnamefont{B.-Q.} \bibnamefont{Ma}},
  \bibinfo{journal}{Journal of Physics G: Nuclear and Particle Physics}
  \textbf{\bibinfo{volume}{34}}, \bibinfo{pages}{1845} (\bibinfo{year}{2007}).

\bibitem[{\citenamefont{Forshaw and Sandapen}(2012)}]{Forshaw:2012im}
\bibinfo{author}{\bibfnamefont{J.~R.} \bibnamefont{Forshaw}} \bibnamefont{and}
  \bibinfo{author}{\bibfnamefont{R.}~\bibnamefont{Sandapen}},
  \bibinfo{journal}{Phys. Rev. Lett.} \textbf{\bibinfo{volume}{109}},
  \bibinfo{pages}{081601} (\bibinfo{year}{2012}), \eprint{1203.6088}.

\bibitem[{\citenamefont{Ahmady and Sandapen}(2013)}]{Ahmady:2013cva}
\bibinfo{author}{\bibfnamefont{M.}~\bibnamefont{Ahmady}} \bibnamefont{and}
  \bibinfo{author}{\bibfnamefont{R.}~\bibnamefont{Sandapen}},
  \bibinfo{journal}{Phys. Rev.} \textbf{\bibinfo{volume}{D88}},
  \bibinfo{pages}{014042} (\bibinfo{year}{2013}), \eprint{1305.1479}.

\bibitem[{\citenamefont{Ahmady et~al.}(2016)\citenamefont{Ahmady, Sandapen, and
  Sharma}}]{Ahmady:2016ujw}
\bibinfo{author}{\bibfnamefont{M.}~\bibnamefont{Ahmady}},
  \bibinfo{author}{\bibfnamefont{R.}~\bibnamefont{Sandapen}}, \bibnamefont{and}
  \bibinfo{author}{\bibfnamefont{N.}~\bibnamefont{Sharma}},
  \bibinfo{journal}{Phys. Rev.} \textbf{\bibinfo{volume}{D94}},
  \bibinfo{pages}{074018} (\bibinfo{year}{2016}), \eprint{1605.07665}.

\bibitem[{\citenamefont{Ahmady et~al.}(2017)\citenamefont{Ahmady, Chishtie, and
  Sandapen}}]{Ahmady:2016ufq}
\bibinfo{author}{\bibfnamefont{M.}~\bibnamefont{Ahmady}},
  \bibinfo{author}{\bibfnamefont{F.}~\bibnamefont{Chishtie}}, \bibnamefont{and}
  \bibinfo{author}{\bibfnamefont{R.}~\bibnamefont{Sandapen}},
  \bibinfo{journal}{Phys. Rev.} \textbf{\bibinfo{volume}{D95}},
  \bibinfo{pages}{074008} (\bibinfo{year}{2017}), \eprint{1609.07024}.

\bibitem[{\citenamefont{Ahmady et~al.}(2018{\natexlab{a}})\citenamefont{Ahmady,
  Mondal, and Sandapen}}]{Ahmady:2018muv}
\bibinfo{author}{\bibfnamefont{M.}~\bibnamefont{Ahmady}},
  \bibinfo{author}{\bibfnamefont{C.}~\bibnamefont{Mondal}}, \bibnamefont{and}
  \bibinfo{author}{\bibfnamefont{R.}~\bibnamefont{Sandapen}},
  \bibinfo{journal}{Phys. Rev.} \textbf{\bibinfo{volume}{D98}},
  \bibinfo{pages}{034010} (\bibinfo{year}{2018}{\natexlab{a}}),
  \eprint{1805.08911}.

\bibitem[{\citenamefont{Ahmady et~al.}(2014)\citenamefont{Ahmady, Lord, and
  Sandapen}}]{Ahmady:2014cpa}
\bibinfo{author}{\bibfnamefont{M.~R.} \bibnamefont{Ahmady}},
  \bibinfo{author}{\bibfnamefont{S.}~\bibnamefont{Lord}}, \bibnamefont{and}
  \bibinfo{author}{\bibfnamefont{R.}~\bibnamefont{Sandapen}},
  \bibinfo{journal}{Phys. Rev.} \textbf{\bibinfo{volume}{D90}},
  \bibinfo{pages}{074010} (\bibinfo{year}{2014}), \eprint{1407.6700}.

\bibitem[{\citenamefont{Ahmady et~al.}(2015)\citenamefont{Ahmady, Hatfield,
  Lord, and Sandapen}}]{Ahmady:2015fha}
\bibinfo{author}{\bibfnamefont{M.}~\bibnamefont{Ahmady}},
  \bibinfo{author}{\bibfnamefont{D.}~\bibnamefont{Hatfield}},
  \bibinfo{author}{\bibfnamefont{S.}~\bibnamefont{Lord}}, \bibnamefont{and}
  \bibinfo{author}{\bibfnamefont{R.}~\bibnamefont{Sandapen}},
  \bibinfo{journal}{Phys. Rev.} \textbf{\bibinfo{volume}{D92}},
  \bibinfo{pages}{114028} (\bibinfo{year}{2015}), \eprint{1508.02327}.

\bibitem[{\citenamefont{Ahmady et~al.}(2018{\natexlab{b}})\citenamefont{Ahmady,
  Leger, Mcintyre, Morrison, and Sandapen}}]{Ahmady:2018fvo}
\bibinfo{author}{\bibfnamefont{M.}~\bibnamefont{Ahmady}},
  \bibinfo{author}{\bibfnamefont{A.}~\bibnamefont{Leger}},
  \bibinfo{author}{\bibfnamefont{Z.}~\bibnamefont{Mcintyre}},
  \bibinfo{author}{\bibfnamefont{A.}~\bibnamefont{Morrison}}, \bibnamefont{and}
  \bibinfo{author}{\bibfnamefont{R.}~\bibnamefont{Sandapen}},
  \bibinfo{journal}{Phys. Rev.} \textbf{\bibinfo{volume}{D98}},
  \bibinfo{pages}{053002} (\bibinfo{year}{2018}{\natexlab{b}}),
  \eprint{1805.02940}.

\bibitem[{\citenamefont{Ahmady et~al.}(2019)\citenamefont{Ahmady, Keller,
  Thibodeau, and Sandapen}}]{Ahmady:2019hag}
\bibinfo{author}{\bibfnamefont{M.}~\bibnamefont{Ahmady}},
  \bibinfo{author}{\bibfnamefont{S.}~\bibnamefont{Keller}},
  \bibinfo{author}{\bibfnamefont{M.}~\bibnamefont{Thibodeau}},
  \bibnamefont{and} \bibinfo{author}{\bibfnamefont{R.}~\bibnamefont{Sandapen}},
  \bibinfo{journal}{Phys. Rev. D} \textbf{\bibinfo{volume}{100}},
  \bibinfo{pages}{113005} (\bibinfo{year}{2019}), \eprint{1910.06829}.

\bibitem[{\citenamefont{Brodsky and de~T\'eramond}(2006)}]{Brodsky:2006uqa}
\bibinfo{author}{\bibfnamefont{S.~J.} \bibnamefont{Brodsky}} \bibnamefont{and}
  \bibinfo{author}{\bibfnamefont{G.~F.} \bibnamefont{de~T\'eramond}},
  \bibinfo{journal}{Phys. Rev. Lett.} \textbf{\bibinfo{volume}{96}},
  \bibinfo{pages}{201601} (\bibinfo{year}{2006}), \eprint{hep-ph/0602252}.

\bibitem[{\citenamefont{de~T\'eramond and Brodsky}(2005)}]{deTeramond:2005su}
\bibinfo{author}{\bibfnamefont{G.~F.} \bibnamefont{de~T\'eramond}}
  \bibnamefont{and} \bibinfo{author}{\bibfnamefont{S.~J.}
  \bibnamefont{Brodsky}}, \bibinfo{journal}{Phys. Rev. Lett.}
  \textbf{\bibinfo{volume}{94}}, \bibinfo{pages}{201601}
  (\bibinfo{year}{2005}), \eprint{hep-th/0501022}.

\bibitem[{\citenamefont{Brodsky et~al.}(2015)\citenamefont{Brodsky,
  de~T\'eramond, Dosch, and Erlich}}]{Brodsky:2014yha}
\bibinfo{author}{\bibfnamefont{S.~J.} \bibnamefont{Brodsky}},
  \bibinfo{author}{\bibfnamefont{G.~F.} \bibnamefont{de~T\'eramond}},
  \bibinfo{author}{\bibfnamefont{H.~G.} \bibnamefont{Dosch}}, \bibnamefont{and}
  \bibinfo{author}{\bibfnamefont{J.}~\bibnamefont{Erlich}},
  \bibinfo{journal}{Phys. Rept.} \textbf{\bibinfo{volume}{584}},
  \bibinfo{pages}{1} (\bibinfo{year}{2015}), \eprint{1407.8131}.

\bibitem[{\citenamefont{de~T\'eramond and Brodsky}(2009)}]{deTeramond:2008ht}
\bibinfo{author}{\bibfnamefont{G.~F.} \bibnamefont{de~T\'eramond}}
  \bibnamefont{and} \bibinfo{author}{\bibfnamefont{S.~J.}
  \bibnamefont{Brodsky}}, \bibinfo{journal}{Phys. Rev. Lett.}
  \textbf{\bibinfo{volume}{102}}, \bibinfo{pages}{081601}
  (\bibinfo{year}{2009}), \eprint{0809.4899}.

\bibitem[{\citenamefont{Brodsky et~al.}(2014)\citenamefont{Brodsky,
  De~Teramond, and Dosch}}]{Brodsky:2013ar}
\bibinfo{author}{\bibfnamefont{S.~J.} \bibnamefont{Brodsky}},
  \bibinfo{author}{\bibfnamefont{G.~F.} \bibnamefont{De~Teramond}},
  \bibnamefont{and} \bibinfo{author}{\bibfnamefont{H.~G.} \bibnamefont{Dosch}},
  \bibinfo{journal}{Phys. Lett. B} \textbf{\bibinfo{volume}{729}},
  \bibinfo{pages}{3} (\bibinfo{year}{2014}), \eprint{1302.4105}.

\bibitem[{\citenamefont{Dosch et~al.}(2015)\citenamefont{Dosch, de~Teramond,
  and Brodsky}}]{Dosch:2015nwa}
\bibinfo{author}{\bibfnamefont{H.~G.} \bibnamefont{Dosch}},
  \bibinfo{author}{\bibfnamefont{G.~F.} \bibnamefont{de~Teramond}},
  \bibnamefont{and} \bibinfo{author}{\bibfnamefont{S.~J.}
  \bibnamefont{Brodsky}}, \bibinfo{journal}{Phys. Rev.}
  \textbf{\bibinfo{volume}{D91}}, \bibinfo{pages}{085016}
  (\bibinfo{year}{2015}), \eprint{1501.00959}.

\bibitem[{\citenamefont{Nielsen and Brodsky}(2018)}]{Nielsen:2018uyn}
\bibinfo{author}{\bibfnamefont{M.}~\bibnamefont{Nielsen}} \bibnamefont{and}
  \bibinfo{author}{\bibfnamefont{S.~J.} \bibnamefont{Brodsky}},
  \bibinfo{journal}{Phys. Rev.} \textbf{\bibinfo{volume}{D97}},
  \bibinfo{pages}{114001} (\bibinfo{year}{2018}), \eprint{1802.09652}.

\bibitem[{\citenamefont{Brodsky et~al.}(2020)\citenamefont{Brodsky,
  de~Teramond, and Dosch}}]{Brodsky:2020ajy}
\bibinfo{author}{\bibfnamefont{S.~J.} \bibnamefont{Brodsky}},
  \bibinfo{author}{\bibfnamefont{G.~F.} \bibnamefont{de~Teramond}},
  \bibnamefont{and} \bibinfo{author}{\bibfnamefont{H.~G.} \bibnamefont{Dosch}}
  (\bibinfo{year}{2020}), \eprint{2004.07756}.

\bibitem[{\citenamefont{Brodsky et~al.}(2016)\citenamefont{Brodsky,
  de~T\'eramond, Dosch, and Lorce}}]{Brodsky:2016rvj}
\bibinfo{author}{\bibfnamefont{S.~J.} \bibnamefont{Brodsky}},
  \bibinfo{author}{\bibfnamefont{G.~F.} \bibnamefont{de~T\'eramond}},
  \bibinfo{author}{\bibfnamefont{H.~G.} \bibnamefont{Dosch}}, \bibnamefont{and}
  \bibinfo{author}{\bibfnamefont{C.}~\bibnamefont{Lorce}},
  \bibinfo{journal}{Int. J. Mod. Phys.} \textbf{\bibinfo{volume}{A31}},
  \bibinfo{pages}{1630029} (\bibinfo{year}{2016}), \eprint{1606.04638}.

\bibitem[{\citenamefont{Brodsky and
  de~T\'eramond}(2008{\natexlab{a}})}]{Brodsky:2008pf}
\bibinfo{author}{\bibfnamefont{S.~J.} \bibnamefont{Brodsky}} \bibnamefont{and}
  \bibinfo{author}{\bibfnamefont{G.~F.} \bibnamefont{de~T\'eramond}},
  \bibinfo{journal}{Phys. Rev.} \textbf{\bibinfo{volume}{D78}},
  \bibinfo{pages}{025032} (\bibinfo{year}{2008}{\natexlab{a}}),
  \eprint{0804.0452}.

\bibitem[{\citenamefont{Drell and Yan}(1970)}]{Drell:1969km}
\bibinfo{author}{\bibfnamefont{S.~D.} \bibnamefont{Drell}} \bibnamefont{and}
  \bibinfo{author}{\bibfnamefont{T.-M.} \bibnamefont{Yan}},
  \bibinfo{journal}{Phys. Rev. Lett.} \textbf{\bibinfo{volume}{24}},
  \bibinfo{pages}{181} (\bibinfo{year}{1970}).

\bibitem[{\citenamefont{West}(1970)}]{West:1970av}
\bibinfo{author}{\bibfnamefont{G.~B.} \bibnamefont{West}},
  \bibinfo{journal}{Phys. Rev. Lett.} \textbf{\bibinfo{volume}{24}},
  \bibinfo{pages}{1206} (\bibinfo{year}{1970}).

\bibitem[{\citenamefont{Brodsky and de~T\'eramond}(2009)}]{Brodsky:2008pg}
\bibinfo{author}{\bibfnamefont{S.~J.} \bibnamefont{Brodsky}} \bibnamefont{and}
  \bibinfo{author}{\bibfnamefont{G.~F.} \bibnamefont{de~T\'eramond}},
  \bibinfo{journal}{Subnucl. Ser.} \textbf{\bibinfo{volume}{45}},
  \bibinfo{pages}{139} (\bibinfo{year}{2009}), \eprint{0802.0514}.

\bibitem[{\citenamefont{Becirevic et~al.}(2003)\citenamefont{Becirevic, Lubicz,
  Mescia, and Tarantino}}]{Becirevic_2003}
\bibinfo{author}{\bibfnamefont{D.}~\bibnamefont{Becirevic}},
  \bibinfo{author}{\bibfnamefont{V.}~\bibnamefont{Lubicz}},
  \bibinfo{author}{\bibfnamefont{F.}~\bibnamefont{Mescia}}, \bibnamefont{and}
  \bibinfo{author}{\bibfnamefont{C.}~\bibnamefont{Tarantino}},
  \bibinfo{journal}{Journal of High Energy Physics}
  \textbf{\bibinfo{volume}{2003}}, \bibinfo{pages}{007} (\bibinfo{year}{2003}).

\bibitem[{\citenamefont{Braun et~al.}(2003)\citenamefont{Braun, Burch,
  Gattringer, G\"ockeler, Lacagnina, Schaefer, and
  Sch\"afer}}]{PhysRevD.68.054501}
\bibinfo{author}{\bibfnamefont{V.~M.} \bibnamefont{Braun}},
  \bibinfo{author}{\bibfnamefont{T.}~\bibnamefont{Burch}},
  \bibinfo{author}{\bibfnamefont{C.}~\bibnamefont{Gattringer}},
  \bibinfo{author}{\bibfnamefont{M.}~\bibnamefont{G\"ockeler}},
  \bibinfo{author}{\bibfnamefont{G.}~\bibnamefont{Lacagnina}},
  \bibinfo{author}{\bibfnamefont{S.}~\bibnamefont{Schaefer}}, \bibnamefont{and}
  \bibinfo{author}{\bibfnamefont{A.}~\bibnamefont{Sch\"afer}}
  (\bibinfo{collaboration}{Bern-Graz-Regensburg Collaboration}),
  \bibinfo{journal}{Phys. Rev. D} \textbf{\bibinfo{volume}{68}},
  \bibinfo{pages}{054501} (\bibinfo{year}{2003}).

\bibitem[{\citenamefont{Chang et~al.}(2017)\citenamefont{Chang, Brodsky, and
  Li}}]{Chang:2016ouf}
\bibinfo{author}{\bibfnamefont{Q.}~\bibnamefont{Chang}},
  \bibinfo{author}{\bibfnamefont{S.~J.} \bibnamefont{Brodsky}},
  \bibnamefont{and} \bibinfo{author}{\bibfnamefont{X.-Q.} \bibnamefont{Li}},
  \bibinfo{journal}{Phys. Rev.} \textbf{\bibinfo{volume}{D95}},
  \bibinfo{pages}{094025} (\bibinfo{year}{2017}), \eprint{1612.05298}.

\bibitem[{\citenamefont{Chang et~al.}(2018)\citenamefont{Chang, Li, Li, and
  Su}}]{Chang:2018aut}
\bibinfo{author}{\bibfnamefont{Q.}~\bibnamefont{Chang}},
  \bibinfo{author}{\bibfnamefont{X.-N.} \bibnamefont{Li}},
  \bibinfo{author}{\bibfnamefont{X.-Q.} \bibnamefont{Li}}, \bibnamefont{and}
  \bibinfo{author}{\bibfnamefont{F.}~\bibnamefont{Su}} (\bibinfo{year}{2018}),
  \eprint{1805.00718}.

\bibitem[{\citenamefont{Brodsky et~al.}(1981)\citenamefont{Brodsky, Huang, and
  Lepage}}]{Brodsky:1981jv}
\bibinfo{author}{\bibfnamefont{S.}~\bibnamefont{Brodsky}},
  \bibinfo{author}{\bibfnamefont{T.}~\bibnamefont{Huang}}, \bibnamefont{and}
  \bibinfo{author}{\bibfnamefont{G.}~\bibnamefont{Lepage}},
  \bibinfo{journal}{Conf. Proc. C} \textbf{\bibinfo{volume}{810816}},
  \bibinfo{pages}{143} (\bibinfo{year}{1981}).

\bibitem[{\citenamefont{Melosh}(1974)}]{PhysRevD.9.1095}
\bibinfo{author}{\bibfnamefont{H.~J.} \bibnamefont{Melosh}},
  \bibinfo{journal}{Phys. Rev. D} \textbf{\bibinfo{volume}{9}},
  \bibinfo{pages}{1095} (\bibinfo{year}{1974}).

\bibitem[{\citenamefont{Huang et~al.}(1994)\citenamefont{Huang, Ma, and
  Shen}}]{Huang:1994dy}
\bibinfo{author}{\bibfnamefont{T.}~\bibnamefont{Huang}},
  \bibinfo{author}{\bibfnamefont{B.-Q.} \bibnamefont{Ma}}, \bibnamefont{and}
  \bibinfo{author}{\bibfnamefont{Q.-X.} \bibnamefont{Shen}},
  \bibinfo{journal}{Phys. Rev. D} \textbf{\bibinfo{volume}{49}},
  \bibinfo{pages}{1490} (\bibinfo{year}{1994}), \eprint{hep-ph/9402285}.

\bibitem[{\citenamefont{Forshaw et~al.}(2004)\citenamefont{Forshaw, Sandapen,
  and Shaw}}]{Forshaw:2003ki}
\bibinfo{author}{\bibfnamefont{J.~R.} \bibnamefont{Forshaw}},
  \bibinfo{author}{\bibfnamefont{R.}~\bibnamefont{Sandapen}}, \bibnamefont{and}
  \bibinfo{author}{\bibfnamefont{G.}~\bibnamefont{Shaw}},
  \bibinfo{journal}{Phys. Rev. D} \textbf{\bibinfo{volume}{69}},
  \bibinfo{pages}{094013} (\bibinfo{year}{2004}), \eprint{hep-ph/0312172}.

\bibitem[{\citenamefont{Choi and Ji}(2005)}]{PhysRevD.72.013004}
\bibinfo{author}{\bibfnamefont{H.-M.} \bibnamefont{Choi}} \bibnamefont{and}
  \bibinfo{author}{\bibfnamefont{C.-R.} \bibnamefont{Ji}},
  \bibinfo{journal}{Phys. Rev. D} \textbf{\bibinfo{volume}{72}},
  \bibinfo{pages}{013004} (\bibinfo{year}{2005}).

\bibitem[{\citenamefont{Melikhov}(1996)}]{Melikhov:1995xz}
\bibinfo{author}{\bibfnamefont{D.}~\bibnamefont{Melikhov}},
  \bibinfo{journal}{Phys. Rev. D} \textbf{\bibinfo{volume}{53}},
  \bibinfo{pages}{2460} (\bibinfo{year}{1996}), \eprint{hep-ph/9509268}.

\bibitem[{\citenamefont{Choi and Ji}(2001)}]{Choi:1999bg}
\bibinfo{author}{\bibfnamefont{H.-M.} \bibnamefont{Choi}} \bibnamefont{and}
  \bibinfo{author}{\bibfnamefont{C.-R.} \bibnamefont{Ji}},
  \bibinfo{journal}{Nucl. Phys. A} \textbf{\bibinfo{volume}{679}},
  \bibinfo{pages}{735} (\bibinfo{year}{2001}), \eprint{hep-ph/9906225}.

\bibitem[{\citenamefont{Qian and Ma}(2008)}]{Qian:2008px}
\bibinfo{author}{\bibfnamefont{W.}~\bibnamefont{Qian}} \bibnamefont{and}
  \bibinfo{author}{\bibfnamefont{B.-Q.} \bibnamefont{Ma}},
  \bibinfo{journal}{Phys. Rev. D} \textbf{\bibinfo{volume}{78}},
  \bibinfo{pages}{074002} (\bibinfo{year}{2008}), \eprint{0809.4411}.

\bibitem[{\citenamefont{Feldmann et~al.}(1998)\citenamefont{Feldmann, Kroll,
  and Stech}}]{Feldmann:1998vh}
\bibinfo{author}{\bibfnamefont{T.}~\bibnamefont{Feldmann}},
  \bibinfo{author}{\bibfnamefont{P.}~\bibnamefont{Kroll}}, \bibnamefont{and}
  \bibinfo{author}{\bibfnamefont{B.}~\bibnamefont{Stech}},
  \bibinfo{journal}{Phys. Rev.} \textbf{\bibinfo{volume}{D58}},
  \bibinfo{pages}{114006} (\bibinfo{year}{1998}), \eprint{hep-ph/9802409}.

\bibitem[{\citenamefont{Brodsky and
  de~T\'eramond}(2008{\natexlab{b}})}]{Brodsky:2007hb}
\bibinfo{author}{\bibfnamefont{S.~J.} \bibnamefont{Brodsky}} \bibnamefont{and}
  \bibinfo{author}{\bibfnamefont{G.~F.} \bibnamefont{de~T\'eramond}},
  \bibinfo{journal}{Phys. Rev.} \textbf{\bibinfo{volume}{D77}},
  \bibinfo{pages}{056007} (\bibinfo{year}{2008}{\natexlab{b}}),
  \eprint{0707.3859}.

\bibitem[{\citenamefont{Brodsky
  et~al.}(2011{\natexlab{a}})\citenamefont{Brodsky, Cao, and
  de~T\'eramond}}]{Brodsky:2011xx}
\bibinfo{author}{\bibfnamefont{S.~J.} \bibnamefont{Brodsky}},
  \bibinfo{author}{\bibfnamefont{F.-G.} \bibnamefont{Cao}}, \bibnamefont{and}
  \bibinfo{author}{\bibfnamefont{G.~F.} \bibnamefont{de~T\'eramond}},
  \bibinfo{journal}{Phys. Rev.} \textbf{\bibinfo{volume}{D84}},
  \bibinfo{pages}{075012} (\bibinfo{year}{2011}{\natexlab{a}}),
  \eprint{1105.3999}.

\bibitem[{\citenamefont{Brodsky
  et~al.}(2011{\natexlab{b}})\citenamefont{Brodsky, Cao, and
  de~T\'eramond}}]{Brodsky:2011yv}
\bibinfo{author}{\bibfnamefont{S.~J.} \bibnamefont{Brodsky}},
  \bibinfo{author}{\bibfnamefont{F.-G.} \bibnamefont{Cao}}, \bibnamefont{and}
  \bibinfo{author}{\bibfnamefont{G.~F.} \bibnamefont{de~T\'eramond}},
  \bibinfo{journal}{Phys. Rev.} \textbf{\bibinfo{volume}{D84}},
  \bibinfo{pages}{033001} (\bibinfo{year}{2011}{\natexlab{b}}),
  \eprint{1104.3364}.

\bibitem[{\citenamefont{Chakrabarti and Mondal}(2013)}]{Chakrabarti:2013dda}
\bibinfo{author}{\bibfnamefont{D.}~\bibnamefont{Chakrabarti}} \bibnamefont{and}
  \bibinfo{author}{\bibfnamefont{C.}~\bibnamefont{Mondal}},
  \bibinfo{journal}{Eur. Phys. J. C} \textbf{\bibinfo{volume}{73}},
  \bibinfo{pages}{2671} (\bibinfo{year}{2013}), \eprint{1307.7995}.

\bibitem[{\citenamefont{Sufian et~al.}(2017)\citenamefont{Sufian, de~Teramond,
  Brodsky, Deur, and Dosch}}]{Sufian:2016hwn}
\bibinfo{author}{\bibfnamefont{R.~S.} \bibnamefont{Sufian}},
  \bibinfo{author}{\bibfnamefont{G.~F.} \bibnamefont{de~Teramond}},
  \bibinfo{author}{\bibfnamefont{S.~J.} \bibnamefont{Brodsky}},
  \bibinfo{author}{\bibfnamefont{A.}~\bibnamefont{Deur}}, \bibnamefont{and}
  \bibinfo{author}{\bibfnamefont{H.~G.} \bibnamefont{Dosch}},
  \bibinfo{journal}{Phys. Rev. D} \textbf{\bibinfo{volume}{95}},
  \bibinfo{pages}{014011} (\bibinfo{year}{2017}), \eprint{1609.06688}.

\bibitem[{\citenamefont{Christ et~al.}(2010)\citenamefont{Christ, Dawson,
  Izubuchi, Jung, Liu, Mawhinney, Sachrajda, Soni, and Zhou}}]{Christ:2010dd}
\bibinfo{author}{\bibfnamefont{N.~H.} \bibnamefont{Christ}},
  \bibinfo{author}{\bibfnamefont{C.}~\bibnamefont{Dawson}},
  \bibinfo{author}{\bibfnamefont{T.}~\bibnamefont{Izubuchi}},
  \bibinfo{author}{\bibfnamefont{C.}~\bibnamefont{Jung}},
  \bibinfo{author}{\bibfnamefont{Q.}~\bibnamefont{Liu}},
  \bibinfo{author}{\bibfnamefont{R.~D.} \bibnamefont{Mawhinney}},
  \bibinfo{author}{\bibfnamefont{C.~T.} \bibnamefont{Sachrajda}},
  \bibinfo{author}{\bibfnamefont{A.}~\bibnamefont{Soni}}, \bibnamefont{and}
  \bibinfo{author}{\bibfnamefont{R.}~\bibnamefont{Zhou}},
  \bibinfo{journal}{Phys. Rev. Lett.} \textbf{\bibinfo{volume}{105}},
  \bibinfo{pages}{241601} (\bibinfo{year}{2010}), \eprint{1002.2999}.

\bibitem[{\citenamefont{Escribano and Frere}(2005)}]{Escribano:2005qq}
\bibinfo{author}{\bibfnamefont{R.}~\bibnamefont{Escribano}} \bibnamefont{and}
  \bibinfo{author}{\bibfnamefont{J.-M.} \bibnamefont{Frere}},
  \bibinfo{journal}{JHEP} \textbf{\bibinfo{volume}{06}}, \bibinfo{pages}{029}
  (\bibinfo{year}{2005}), \eprint{hep-ph/0501072}.

\bibitem[{\citenamefont{Munz et~al.}(1995)\citenamefont{Munz, Resag, Metsch,
  and Petry}}]{Munz:1994ty}
\bibinfo{author}{\bibfnamefont{C.}~\bibnamefont{Munz}},
  \bibinfo{author}{\bibfnamefont{J.}~\bibnamefont{Resag}},
  \bibinfo{author}{\bibfnamefont{B.}~\bibnamefont{Metsch}}, \bibnamefont{and}
  \bibinfo{author}{\bibfnamefont{H.}~\bibnamefont{Petry}},
  \bibinfo{journal}{Phys. Rev. C} \textbf{\bibinfo{volume}{52}},
  \bibinfo{pages}{2110} (\bibinfo{year}{1995}), \eprint{nucl-th/9406035}.

\end{thebibliography}

\end{document}